\documentclass[fleqn,usenatbib]{mnras}

\usepackage{newtxtext,newtxmath}

\usepackage[T1]{fontenc}

\DeclareRobustCommand{\VAN}[3]{#2}
\let\VANthebibliography\thebibliography
\def\thebibliography{\DeclareRobustCommand{\VAN}[3]{##3}\VANthebibliography}

\usepackage{graphicx}	
\usepackage{amsmath}	

\title[Dynamics of z $\sim 4.5$ dusty star-forming galaxies]{Dynamical properties of z $\sim 4.5$ dusty star-forming galaxies and their connection with local early type galaxies}

\author[F. Rizzo et al.]{
Francesca Rizzo$^{1, 2, 3},$\thanks{E-mail: francesca.rizzo@nbi.ku.dk}
Simona Vegetti$^{1}$,
Filippo Fraternali$^{4}$,
Hannah R. Stacey$^{1}$,
Devon Powell$^{1}$
\\
$^{1}$Max-Planck Institute for Astrophysics, Karl-Schwarzschild Str. 1, D-85748, Garching, Germany\\
$^{2}$Cosmic Dawn Center (DAWN), Denmark\\
$^{3}$Niels Bohr Institute, University of Copenhagen, Lyngbyvej 2, DK-2100 Copenhagen Ø, Denmark\\
$^{4}$University of Groningen, Kapteyn Astronomical Institute, Postbus 800, 9700 AV Groningen, The Netherlands
}

\date{Accepted 2021 August 3. Received 2021 August 3; in original form 2021 February 10}

\pubyear{2021}

\begin{document}
\label{firstpage}
\pagerange{\pageref{firstpage}--\pageref{lastpage}}
\maketitle

\begin{abstract}
There is a large consensus that gas in high-$z$ galaxies is highly turbulent, because of a combination of stellar feedback processes and gravitational instabilities driven by mergers and gas accretion. In this paper, we present the analysis of a sample of five Dusty Star Forming Galaxies (DSFGs) at $4 \lesssim z\lesssim 5$. Taking advantage of the magnifying power of strong gravitational lensing, we quantified their kinematic and dynamical properties from ALMA observations of their [CII] emission line. We combined the dynamical measurements obtained for these galaxies with those obtained from previous studies to build the largest sample of $z \sim 4.5$ galaxies with high-quality data and sub-kpc spatial resolutions, so far. We found that all galaxies in the sample are dynamically cold, with rotation-to-random motion ratios, $V/\sigma$, between 7 to 15. The relation between their velocity dispersions and their star-formation rates indicates that stellar feedback is sufficient to sustain the turbulence within these galaxies and no further mechanisms are needed. In addition, we performed a rotation curve decomposition to infer the relative contribution of the baryonic (gas, stars) and dark matter components to the total gravitational potentials. This analysis allowed us to compare the structural properties of the studied DSFGs with those of their descendants, the local early type galaxies. In particular, we found that five out of six galaxies of the sample show the dynamical signature of a bulge, indicating that the spheroidal component is already in place at $z \sim 4.5$.
\end{abstract}




\section{Introduction}
Within the framework of current galaxy formation and evolutionary models, galaxies grow by acquiring material through gas accretion and mergers \citep[e.g.,][]{dekel09, rodriguez, naab17}. Feedback processes driven by active galactic nuclei (AGN) or star formation are able to temper the growth of galaxies through the heating or expulsion of gas \citep{hopkins, silk, nelson}. On the other hand, many of these feedback processes operate on physical scales that are well below the resolution of current cosmological simulations, and are, therefore, usually parametrized using simple sub-grid prescriptions which are mostly calibrated to match observations in the local Universe \citep[e.g.,][]{crain, kim, vogelsberger}.\\
Observational evidence able to give a consistent and quantitative picture of how galaxies grow and evolve across cosmic time is still lacking. For example, the importance of mergers in driving the stellar mass growth, as well as in determining the resulting kinematic and chemical properties of galaxies, is still a matter of debate \citep[e.g.][]{oesch, satyapal, eliche}. 
The influence of feedback processes (e.g., outflows, turbulence), mergers and gas accretion in regulating the growth of stellar mass is observationally challenging to evaluate, even in the local Universe, since complex physical mechanisms operating at different scales need to be identified and constrained \citep[e.g.,][]{cicone, concas, mcquinn}. 
Moreover, at high redshift, these challanges are exacerbated due to the limited angular resolution and signal-to-noise ratio (SNR) of the current observations \citep[e.g.,][]{chisholm, stanley, ginolfi}. One of the ways to partly overcome these observational limitations is to target strongly gravitationally lensed galaxies. The magnification provided by gravitational lensing increases the solid angle of background sources and hence their observed total flux \citep{schneider}. As a result, distant objects can be observed with either increased spatial resolution (for spatially-resolved observations) or increased SNR (for unresolved observations).

Recently, \citet{rizzo3} showed the power of high resolution observations in unveiling the dynamical properties of a lensed dusty star-forming galaxy (DSFG), SPT0418-47, at $z = 4.2$. In particular, they found that SPT0418-47 has dynamical properties similar to those of local spiral galaxies: it is rotationally supported and has a low level of turbulence, that is, it is dynamically cold. These features indicate that, unexpectedly, the high star-formation rate (SFR) and the gas fraction measured for this DSFG do not drive high turbulent motions nor affect the stability of the galaxy. \citet{fraternali20} obtained a similar result for two non-lensed DSFGs at $z \approx 4.5$. 
The global SFR and interstellar medium (ISM) properties of these three galaxies are typical of the population of DSFGs \citep{hodge, aravena, gomez}, suggesting that such systems are common among star-forming galaxies at high redshifts. However, reproducing the existence of galaxies with such a large rotational velocity, SFR and content of cold gas remains challenging for most numerical simulations and semi-analytic models \citep[e.g.][]{dekel, zolotov15, pillepich, dekel20}. Robust observational constraints on the spatially-resolved kinematic properties of star-forming galaxies at $z \gtrsim 4$ are thus necessary to inform us about how galaxies build up their mass and how they are affected by different physical processes.

\citet{hodge12} obtained the first spatially-resolved kinematics of a galaxy at $z = 4$, GN20, using CO observations from the Very Large Array. However, the long integration times (e.g., $\approx$ 120 hours for GN20) needed to spatially resolve the faint CO transitions made very challenging the extension of this analysis to a large sample of $z \gtrsim 4$ galaxies. In the past six years, the advent of the Atacama Large Millimitre/Submillimitre Array \citep[ALMA,][]{wootten} has opened up a new frontier for detailed kinematic studies of galaxies in the redshift range $z \sim 4$ - 6 using the $^2P_{3/2} \rightarrow ^{2}P_{1/2}$ transition at 1900.5469 GHz (157.74 $\mu$m, [CII]) of the ionized carbon C+ \citep[e.g.,][]{debreuck14, jones17, smit, neelman, rizzo3, fraternali20}. The 158 $\mu$m [CII] emission line is, in fact, a powerful tool to investigate the gas physical conditions in the distant Universe: it is typically the brightest fine-structure line emitted in star-forming galaxies, representing $\sim0.1$ - $1$ percent of the total far-infrared luminosity in the most active systems \citep{stacey10, sargsyan}. 

By studying the structural properties of star-forming galaxies at $4 \lesssim z\lesssim 5$, we also gain insights into the formation of massive quiescent galaxies at lower redshift. Stellar population studies of local early type galaxies (ETGs) have shown, indeed, that more than half of the stars in the most massive galaxies were formed at $z \gtrsim$ 3 \citep{mcdermid}. Furthermore, recent studies have found a spectroscopically-confirmed population of massive, quiescent galaxies already at $3 \lesssim z \lesssim 4$ \citep{glazebrook, tanaka, valentino}. The properties of the quiescent populations both at low and intermediate redshifts suggest that their progenitors should belong to the star-forming population at $z \gtrsim$ 3 \citep{mcdermid, belli, hodge20}. The matching number densities, stellar masses and sizes of the massive star-forming at $z \gtrsim 4$ and quiescent populations at lower redshifts seem to support this picture \citep{toft, valentino}.
The common structural feature of these quiescent systems is the presence of a spheroidal component, i.e., a bulge \citep[e.g.,][]{kraj, lang, vanderwel, belli, lustig}. Despite several theories and scenarios have been proposed \citep[e.g.][]{elemgreen08, bournaud09, kormendy, erwin, brooks}, to this date, it is still unclear which are the main mechanisms driving the rapid formation of these bulges already at these early times \citep[e.g.][]{kormendy, dimauro, nelsone, tacchella18}.
A robust measurement of the kinematic and structural properties of $4 \lesssim z\lesssim 5$ galaxies requires high spatial resolution observations, which nowadays can be achieved efficiently by targeting strongly gravitationally lensed galaxies with ALMA. In this paper, we extend the dynamical analysis of SPT0418-47 \citep{rizzo3} to a sample of five lensed DSFGs. Using the 3D kinematic-lens modelling technique developed by \citet{rizzo2} and applied to ALMA interferometric data, we reconstruct the dust and [CII] emission in these galaxies on sub-kpc scales. In Section \ref{sec:data}, we describe the targets, observations and data reduction process. In Section \ref{sec:analysis}, we summarise our lens modelling and source reconstruction technique and the dynamical analysis on the background sources. In Sections \ref{sec:result} and \ref{sec:discussion}, we present the results and discuss the implications of our findings. The main conclusions of this work are summarized in Section \ref{sec:conclusions}. Throughout the paper, we assume a $\Lambda$CDM cosmology, with Hubble constant $H_0$ = 67.8 km/s/Mpc, matter density $\Omega_{\mathrm{m}}$ = 0.308, and vacuum energy density $\Omega_{\mathrm{\Lambda}}$ = 0.691 from \citet{planck}. 

\section{Sample and observations}
\label{sec:data}

\subsection{The 158-$\mu$m [C II] emission line: a tracer of the gas kinematics at high-$z$}
The 158-$\mu$m [CII] emission line can trace multiple phases of the ISM, including the warm ionized, the warm and cold neutral atomic, and the dense molecular medium \citep{stacey}, due to the lower ionization potential of 11.3 eV of the atomic carbon with respect to HI. However, several studies \citep{rigopoulou, cormier, debreuck19} found that more than 60 percent of the [CII] emission originates in the photodissociation regions, the external layers of molecular clouds heated by the far-ultraviolet photons emitted from OB stars. In these regions, both atomic and molecular hydrogen, as well as electrons, can collisionally excite the ground state of C+ ions producing the [CII] emission line. 
Furthermore, the molecular gas can be an efficient emitter of [CII] rather than CO \citep{pineda, nordon, glover}. Specifically, in low-metallicity environments or the outer regions of molecular clouds, H$_2$ self-shields and survives while CO can be easily photodissociated into C and C+ \citep[e.g.,][]{papa, wolfire, madden}. This wide range of physical conditions makes [CII] an excellent tracer of the kinematics of high-$z$ star-forming galaxies over large areas of their discs.
\subsection{Selection of the sample}
The targets in this paper are five gravitationally lensed DSFGs (Table \ref{tab:sample}).
We collect the sample by selecting from the ALMA public archive star-forming galaxies at $4 \lesssim z\lesssim 5$ which are lensed by a galaxy or a galaxy group and with data of angular ($\lesssim 0.3$ arcsec) and spectral ($\lesssim$ 40 km/s) resolutions high enough to resolve the emission of the 158-$\mu$m [CII] emission line (Table \ref{tab:sample}). In addition, we select observations with a median (over the spectral channels) SNR of the order of 10. The modelling code that we will apply to these data requires, indeed, fairly high SNRs \citep{rizzo2}, since it fits the data in all spectral channels, without requiring any projections to first and second moment maps. This approach is fundamental to obtain robust kinematic measurements, not affected by the so-called beam smearing effect, which may cause strong bias to low values of the rotation velocities and high values of the velocity dispersions \citep{edt1}. Furthermore, these high SNRs guarantee that no spatial binning is required, so that we can fully take advantage of the angular resolution of ALMA and probe the kinematics of our sample on sub-kpc scales.\\
The five lensing system of our selected sample were identified in the South Pole Telescope (SPT) survey \citep{carlstrom, vieira10, vieira, reuter} and have existing ancillary spectroscopic and imaging data in the sub-mm and far-infrared wavelength range, from which the redshifts of the background galaxies (Table \ref{tab:sample}) and the infrared luminosities (see Section \ref{sec:continuum}) were derived \citep[e.g.,][]{weis, strandet, reuter, aravena}. 
\subsection{ALMA data}
The observations for each target have two spectral windows covering the redshifted rest frequency of the [CII] line and two spectral windows for the continuum. Each spectral window has 240 channels and a 1.875 GHz bandwidth. In this paper, we make use of the calibrated measurement sets provided by the European ALMA Regional Centre \citep{hatz}, that calibrated the raw visibility data using the ALMA pipeline in the {\sc CASA} package \citep{mcmullin}. These data were then inspected to confirm the quality of the pipeline calibration and that no further flagging was required. We then performed one or two rounds of phase-only self-calibration on the continuum data, using solution
intervals of the scan length or half the scan length and applied the complex gain corrections from the continuum to the spectral windows containing the line. The continuum is subtracted from the line spectral window using {\sc UVCONTSUB}. The data are averaged into groups of between four (e.g., SPT0345-47, SPT2132-58) and six (e.g., SPT0441-46, SPT2146-55) channels in order to increase the overall SNR.
This procedure results in channels with a typical velocity width of $\approx 30$ km s$^{-1}$ (see Table \ref{tab:sample}). \\
The targets are imaged with natural weighting of the visibilities and deconvolved using {\sc CLEAN} \citep{hog}. In the panels a to c of Fig.~ \ref{fig:01} and  Figs.~\ref{fig:ap4}-\ref{fig:ap13} in Appendix \ref{appendixb}, we show the spectral line moment maps of the lensed galaxies.
We note that these images are intended only for visualisation, as the analysis is performed on the visibility data directly (see Section \ref{sec:analysis}).

\begin{table*}
	\centering
	\caption[]{Summary of the observed targets. Columns one and two: IAU and short name. Column three: lens redshifts from \citet{spilker}, when available. Column four: source redshifts from \citet{reuter}. Column five: total on-source integration times. Column six: ALMA project codes (PI: K. Litke). Column seven: beam size of the [CII] observations. Column eight: channel width of the [CII] data cube.}
	\label{tab:sample}
	\begin{tabular}{cccccccc} 
		\hline
		\noalign{\smallskip}
		IAU Name & Short Name & $z_\mathrm{lens}$ & $z_\mathrm{source}$ & $t_\mathrm{s}$ & Project code & Beam size & Channel width\\
				 &  &  &  & min &  & arcsec $\times$ arcsec & km s$^{-1}$\\
	    \noalign{\smallskip}
		\hline
		\noalign{\smallskip}
		 SPT-S J011308-4617.7 & SPT0113-46 & -- & 4.23 & 13.9 & 2016.1.01499.S & 0.35 $\times$ 0.19 & 32.2\\\noalign{\vspace{1pt}}
		 SPT-S J034510-4725.7 & SPT0345-47 & 0.36 & 4.29 & 29.2 & 2016.1.01499.S & 0.18 $\times$ 0.16 & 26.1\\\noalign{\vspace{1pt}}
		 SPT-S J044143-4605.5 & SPT0441-46 & 0.88 & 4.48 & 28.0 & 2016.1.01499.S &  0.23 $\times$ 0.19 & 40.5\\\noalign{\vspace{1pt}}
		 SPT-S J214654-5507.9 & SPT2146-55 & -- & 4.57 & 23.8 & 2016.1.01499.S & 0.23 $\times$ 0.20 & 41.1\\\noalign{\vspace{1pt}}
		SPT-S J213244-5803.1 & SPT2132-58 & -- & 4.77 & 24.4 & 2016.1.01499.S &  0.25 $\times$ 0.22 & 28.4\\\noalign{\vspace{1pt}}
    \noalign{\smallskip}
    \hline
    \noalign{\medskip}
	\end{tabular}
\end{table*}
\section{Analysis}
\label{sec:analysis}
In this section, we provide an overview of the lens-kinematic modelling technique with which we derive the [CII] surface brightness distribution in each spectral channel, as well as the kinematic and dynamical properties of the lensed galaxy. In addition, we also recover the surface brightness distribution of the sources from the continuum dataset.\\
To infer the lens mass models, the sources and their kinematic properties, we make use of the Bayesian pixellated technique developed by \citet{rizzo2} and further extended to the visibility domain by \citet{powell}. We refer the readers to the above papers for a detailed description of the methodology. Here, we provide a summary and emphasise that our modelling technique enables us to simultaneously reconstruct the lensing mass distribution (see Section \ref{sec:lens_model}) and the kinematics of the source (see Section \ref{sec:src_kin}) from the same three-dimensional (3D, two spatial and one spectral dimension) data, by fitting directly in their native visibility plane. The dynamical analysis of the background galaxies is described in Section \ref{sec:src_dyn}.\\

\subsection{Lens mass modelling}
\label{sec:lens_model}
We assume that the lens mass distribution is described by an elliptical power-law profile with an external shear component. The parameters defining the power-law profile are the surface mass density normalization $\kappa_0$, the position angle $\theta$, the axis ratio $q$ and the slope $\gamma$. The lens mass model parameters define the projected mass density distribution, normalised to the critical density, as follows
\begin{equation}
\kappa\left(x,y\right)=\frac{\kappa_0\left(2-\frac{\gamma}{2}\right)q^{\gamma-\frac{3}{2}}}{2\left[q^2\left(x^2+r_{\mathrm{c}}^2\right)+y^2\right]^{\frac{\gamma-1}{2}}}\,.
\label{eq:mass_lens}
\end{equation}
The parameters defining the shear component are its strength $\Gamma_{\mathrm{sh}}$ and position angle $\theta_{\mathrm{sh}}$.

\subsection{Source reconstruction and kinematic properties}
\label{sec:src_kin}
In our methodology, the source is pixellated, that is, its surface brightness distribution is reconstructed on a grid that is adaptive to the varying resolution given by the local lensing magnification \citep{vegetti09}. For the sample studied in this paper, the minimum spatial resolutions range from $\approx 20$ to $130$ pc and the median spatial resolutions range from $\approx 170$ to $300$ pc (see Appendix \ref{appendixc} for further details).\\
The kinematic-lens modelling technique adopted in this paper \citep{rizzo2} makes use of a 3D kinematic model, that describes a rotating disc, as a regularising hyperprior to the pixellated source reconstruction. The kinematic model is defined by the geometrical parameters (inclination $i$ and position angle $PA$) and the parameters describing the rotation and velocity dispersion curves. In particular, for all but one of the sources in the sample, we adopted a multi-parameter function \citep{rix} for their rotation curve
\begin{equation}
V_{\mathrm{rot}}\left(R\right) = V_{\mathrm{t}}\frac{\left(1+\frac{R_{\mathrm{t}}}{R}\right)^{\beta}}{\left[1+\left(\frac{R_{\mathrm{t}}}{R}\right)^{\xi}\right]^{1/\xi}},
\label{eq:mult}
\end{equation}
since it is flexible enough to reproduce a large variety of rotation curves \citep{rizzo2}. In contrast, we found that for SPT2132-58 a simpler arctangent function \citep[e.g.,][]{swinbank11, harrison, turner},
\begin{equation}
V_{\mathrm{rot}}\left(R\right) = \frac{2}{\pi}\,V_{\mathrm{t}}\,\arctan\left(\frac{R}{R_{\mathrm{t}}}\right),
\label{eq:acrtg}
\end{equation}
was sufficient to fit the data to the noise level (see Section \ref{sec:result:kin}). We believe this to be related to the small Einstein ring $\approx 0.3$ arcsec of this system that results in a lower spatial resolution and therefore in fewer available constraints. \\
Measurements of the cold gas velocity dispersion in nearby and high-$z$ galaxies show that for most galaxies the values of $\sigma$ decrease with increasing galactocentric radius, following an exponential or a linear trend \citep[e.g.,][]{tamburro, mogotsi, bacchini19, sharda, lelli21, bacchini}. For all of the galaxies in the sample, we assume the velocity dispersion curve to have an exponential profile
\begin{equation}
\sigma\left(R\right) = \sigma_{0}\,e^{-\frac{R}{R_{\sigma}}}.
\label{eq:exp}
\end{equation}

\subsubsection{Modelling strategy}\label{sec:mod_strategy}
To infer the lens and kinematic parameters, and the source we apply an optimisation scheme, which is divided in the following stages:
\begin{enumerate}
    \item To find an initial guess for the lens parameters, we start by modelling the zeroth-moment map of the data. \label{step:1}
    \item  We then model the 3D data cube and reoptimize for the lens parameters, starting from the values found in step \ref{step:1}. At this stage, the source is reconstructed on a pixellated grid, without the kinematic hyperprior.\label{step:2}
    \item We fix the lens parameters to the values found in step \ref{step:2} and we activate the source hyperprior, optimizing for the kinematic parameters. We check that the reconstructed source obtained at this stage is consistent with that obtained at point \ref{step:2}. 
    In Appendix \ref{appendix:test}, Figs.~\ref{fig:spt0346_ch} and \ref{fig:spt0346_pv}, we show an example of the reconstruction of a lensed source which is not well described by a rotating disc.  \label{step:3}  
\end{enumerate}

\subsection{Rotation curve decomposition} \label{sec:src_dyn}
To quantify how the different matter components contribute to the total galactic gravitational potential of the source galaxies, we perform a rotation curve decomposition, similarly to what has been done in \citet{rizzo3}. The rotation velocity $V_{\mathrm{rot}}(R)$ of the gas is, indeed, related to the total gravitation potential of the galaxy $\Phi$\footnote{This relation is obtained under the assumption that $\Phi$ is axisymmetric} by
\begin{equation}
R \left(\frac{\partial \Phi}{\partial R}\right)_{z=0} = V_{\mathrm{c}}^2 =  V_{\mathrm{rot}}^2 + V_{\mathrm{A}}^2,
\end{equation}
where $V_{\mathrm{c}}$ is the circular velocity, and $V_{\mathrm{A}}$ is the asymmetric-drift correction that accounts for the pressure support due to the random motions \citep[see equation 6 in][]{rizzo3}. To derive $V_{\mathrm{A}}$ not only the velocity dispersion profile $\sigma(R)$ is needed, but also the spatial distribution of the gas component, traced by the [CII] emission line. To this end, we assume that the gas has a spatial distribution described by an exponential profile,
\begin{equation}
\Sigma_{\mathrm{gas}} = \Sigma_{0} \exp{ \left(-\frac{R}{R_{\mathrm{gas}}}\right)}.
\end{equation}
To measure the scale radius $R_{\mathrm{gas}}$, we divide the zeroth-moment map of the reconstructed sources (Fig.~\ref{fig:01} and Figs. \ref{fig:ap4} to \ref{fig:ap13}, panel d) in rings with centers, $PA$ and $i$ defined by the values of the kinematic models. We then azimuthally average the values inside each ring to obtain the surface density profile that is then fitted with the exponential profile (Fig.~\ref{fig:exp}).\\

\noindent We model the circular velocity as 
\begin{equation}
V_{\mathrm{c}} = \sqrt{V^2_{\mathrm{star}} +V^2_{\mathrm{gas}}+ V^2_{\mathrm{DM}}}, 
\label{eq:vc}
\end{equation}
where $V_{\mathrm{star}}$, $V_{\mathrm{gas}}$, $V_{\mathrm{DM}}$ are the contributions of the stellar, gas and dark matter components to the circular speed for which we make the following assumptions:
\begin{itemize}
\item the stellar component is described by a Sérsic profile \citep{limaneto, terzic}. $V_{\mathrm{star}}$ is, therefore, 
\begin{equation}
V_{\mathrm{star}} = \sqrt{\frac{G M_{\mathrm{star}}}{R} \frac{\gamma(n(3-p), b(R/R_{\mathrm{e}})^{1/n})}{\Gamma(n(3-p))}}
\label{eq:vstar}
\end{equation}
where $M_{\mathrm{star}}$ is the total stellar mass, $R_{\mathrm{e}}$ is the effective radius and $n$ is the Sérsic index. In equation (\ref{eq:vstar}), $\gamma$ and $\Gamma$ are the incomplete and complete gamma function, respectively, while the parameters $p$ and $b$ are functions of the Sérsic index $n$ \citep[see Section 2 in][]{terzic}. The lack of spatially resolved data from the rest-frame optical or near-infraed emission prevents the fitting of two stellar components (i.e., bulge and disc), due to the strong degeneracies between the two. The single Sérsic component employed in the dynamical fitting should be, therefore, considered as a global description of the stellar distribution.
\item The gas component has the same distribution as the [CII] emission line and it is defined by an exponential profile so that
\begin{equation}
V_{\mathrm{gas}} = \sqrt{\frac{2 G M_{\mathrm{gas}}}{R_{\mathrm{gas}}} y^2 [I_0(y)K_0(y)-I_1(y)K_1(y)]}
\label{eq:vgas}
\end{equation}
where $y = R/R_{\mathrm{gas}}$ and $K_0, K_1, I_0, I_1$ are the modified Bessel functions \citep{binney}. Since the scale length $R_{\mathrm{gas}}$ is fixed at the value found above, the only free parameter of the fit for $V_{\mathrm{gas}}$ is the conversion factor ($\alpha_{\mathrm{[CII]}}$) between the total [CII] luminosity and the total gas mass, $M_{\mathrm{gas}} = \alpha_{\mathrm{[CII]}} L_{\mathrm{[CII]}} $ \citep[see][and Appendix \ref{appendixa} for further details]{rizzo3}. 
\item With the current data, we can probe only the inner regions of our galaxy sample, mostly dominated by the baryonic components. However, for self-consistency, we make simple but reasonable assumptions for taking into account the dark matter contribution to the gravitational potential. The dark matter halo is modelled as a Navarro-Frenk-White \citep[NFW][]{navarro} spherical halo,
\begin{equation}
V_{\mathrm{DM}} = \sqrt{\frac{G M_{\mathrm{DM}}}{R_{\mathrm{DM}}}\frac{\ln(1+cx)-cx/(1+cx)}{x[\ln(1+c)-c/(1+c)]}},
\label{eq:vdm}
\end{equation}
where $c$ is the concentration parameters and $M_{\mathrm{DM}}$ and $R_{\mathrm{DM}}$ are the virial mass and radius, respectively. To reduce the number of free parameters, we fixed the value of $c$ to the value obtained by averaging the values of the concentration parameters for dark matter halo masses in the mass range between $10^{10} M_{\odot}$ and $10^{13} M_{\odot}$ and assuming the mass-concentration relation estimated in N-body cosmological simulations \citep[][see Table \ref{tab:c}]{dutton}. We notice that at these redshifts, the concentration is almost independent of the dark-matter halo mass varying by at most 8 percent for a variation of 3 orders of magnitude in the halo mass \citep[see][for further details on this assumption]{rizzo3}. The only free parameter of $V_{\mathrm{DM}}$ is the virial mass $M_{\mathrm{DM}}$, since the virial radius $R_{\mathrm{DM}}$ can be expressed as a function of $M_{\mathrm{DM}}$.
\end{itemize}
To fit the circular velocities and compute the Bayesian posterior distribution of the free parameters ($M_{\mathrm{star}}$, $R_{\mathrm{e}}$, $n$, $\alpha_{\mathrm{[CII]}}$, $M_{\mathrm{DM}}$) we use {\sc DYNESTY}, a python implementation of the Dynamic Nested Sampling algorithm \citep{speagle}. We use flat prior distributions for $R_{\mathrm{e}}$, $n$ and $\alpha_{\mathrm{[CII]}}$ and log-uniform priors for the mass parameters (see Table \ref{tab:prior} in Appendix \ref{appendixa}).

\section{Results}
\label{sec:result}
\begin{figure*}
    \centering
    \includegraphics[width=2\columnwidth]{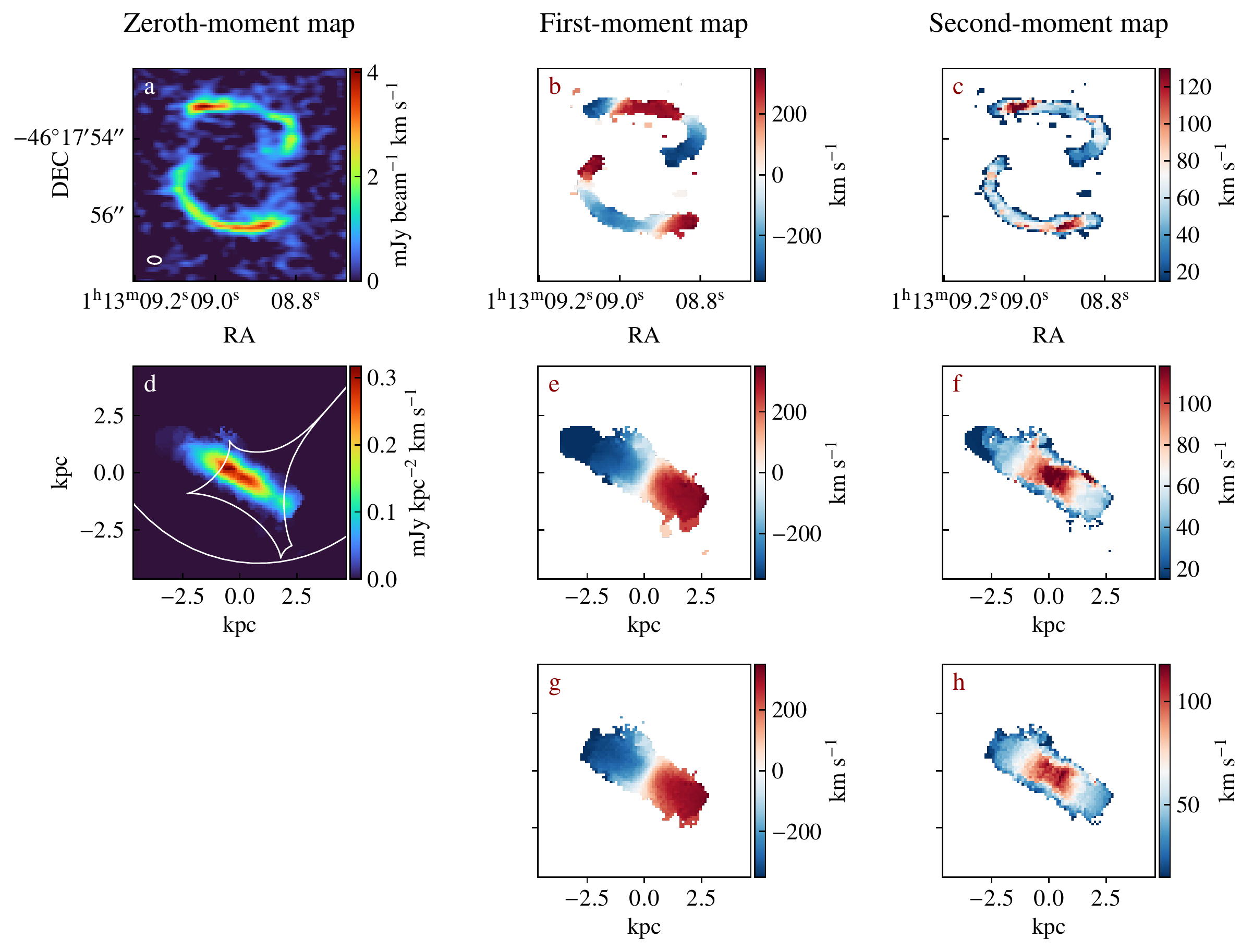}
    \caption[Moment maps for SPT0113-46]{Moment maps for SPT0113-46. Panels a, b and c: the observed [CII] zeroth-, first- and second-moment maps. The beam size, shown as a white ellipse on the lower left corner of panel a, is $0.35 \times 0.19$ arcsec$^{2}$ at a position angle of 87.0$^{\circ}$. Panels d, e and f: zeroth-, first- and second-moment maps of the reconstructed source. The white curves in panel d show the caustics. Panels g and h: first- and second-moment maps of the kinematic model. These maps are intended only for visualisation as the full analysis is performed on the data cube.}
    \label{fig:01}
\end{figure*}

\begin{figure*}
    \centering
    \includegraphics[width=0.95\textwidth]{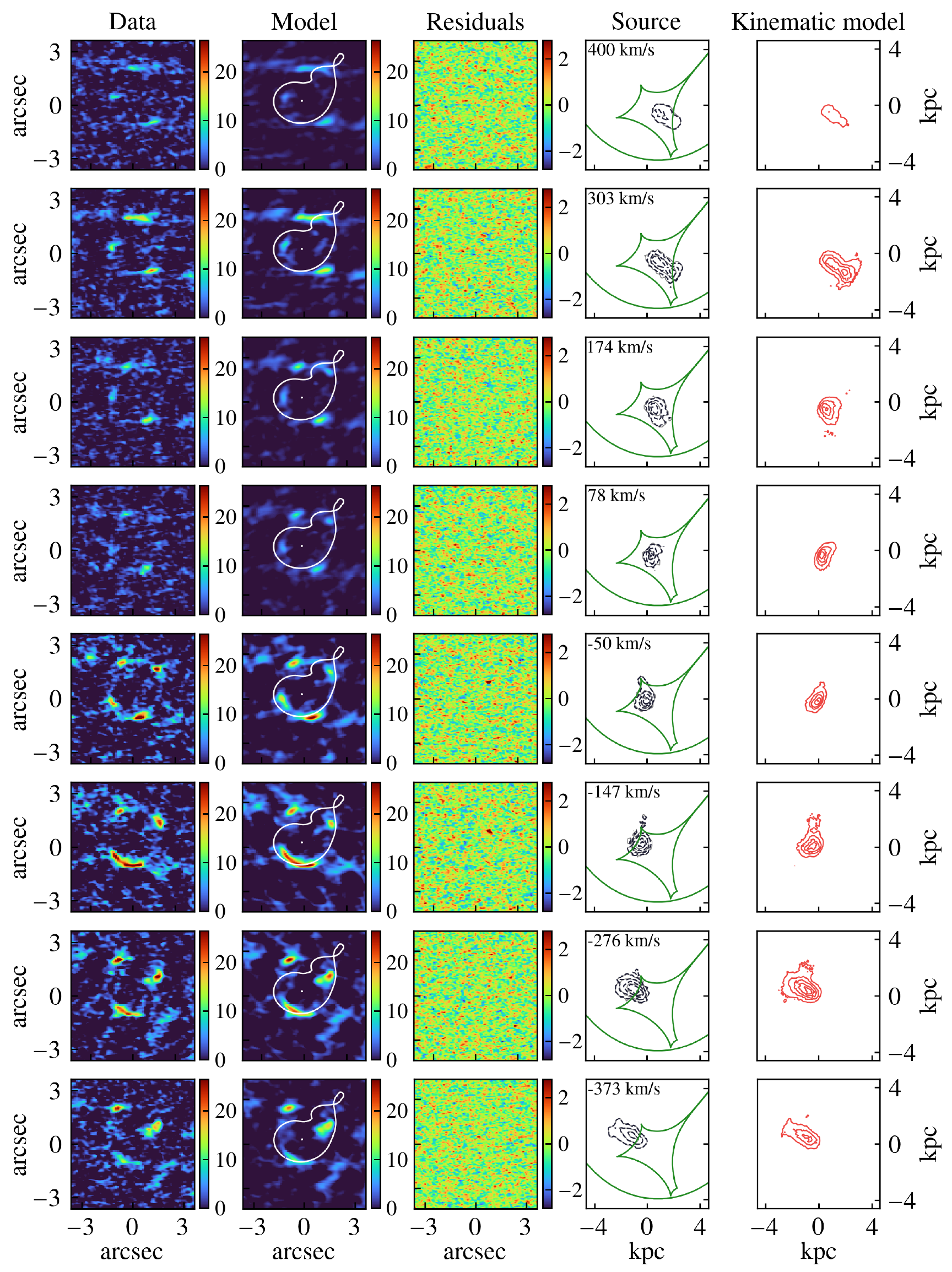}
    \caption{Representative channel maps for SPT0113-46 at the velocity shown on the upper left corner of column 4. Columns 1 and 2: dirty image of the data and the model. The colourbar units are mJy/beam. The white curves in the model columns show the critical curves. Column 3: dirty image residuals (data - model) normalized to the noise. Columns 4 and 5: reconstructed source and kinematic model with contours set at $n$ = [0.1, 0.3, 0.5, 0.7, 0.9] times the value of the maximum flux of the reconstructed source. In the source columns: the dashed black and solid gray contours show the sources obtained with and without a rotating disk as a hyperprior for the source reconstruction, respectively. The green curves in the source columns show the caustics.}
    \label{fig:03}
\end{figure*}

\begin{figure*}
    \centering
    \includegraphics[width=0.8\textwidth]{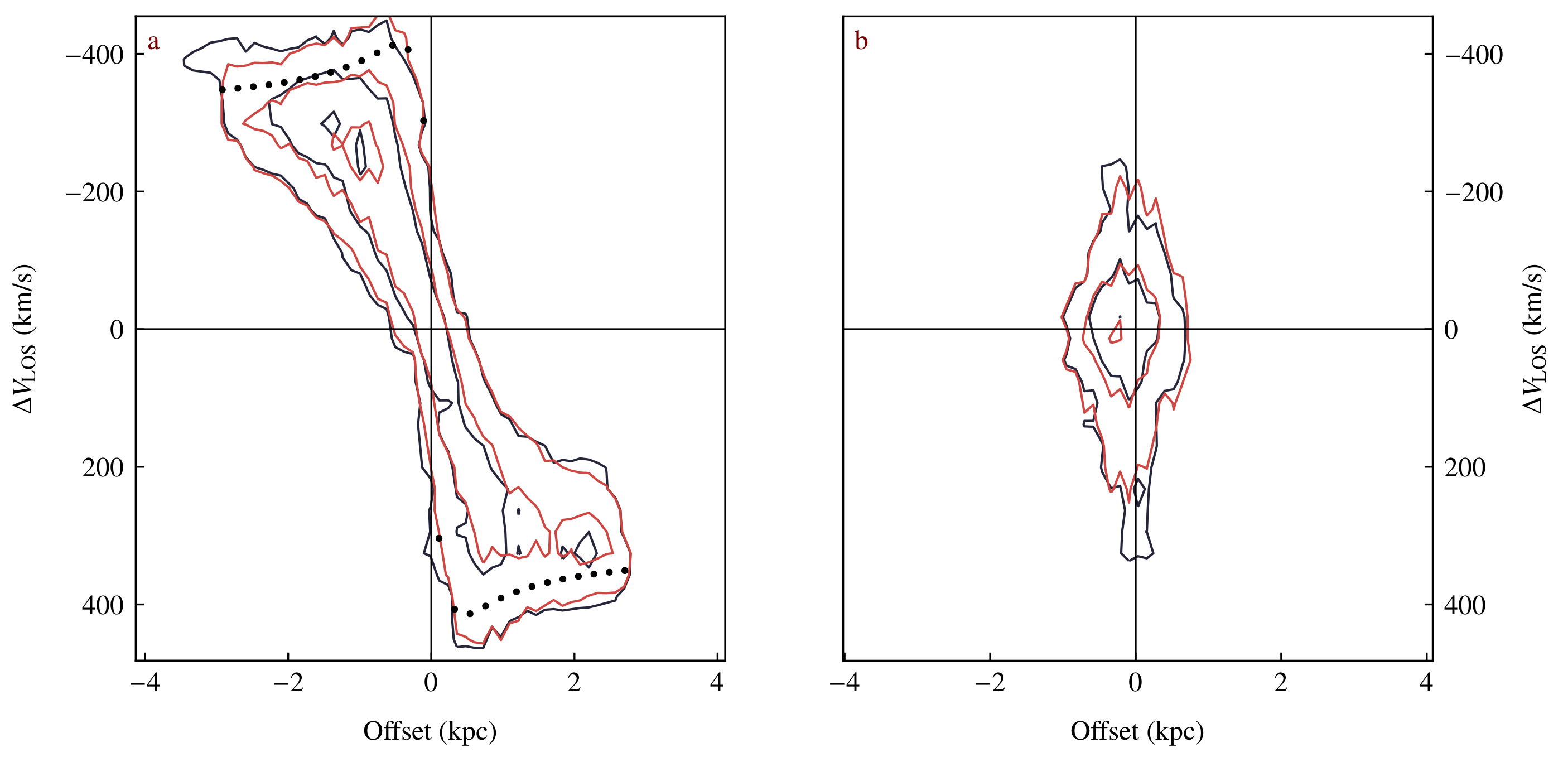}
    \caption[Position-velocity diagrams for SPT0113-46]{Position-velocity diagrams for SPT0113-46. The x-axes show the offset along the major (panel a) and minor axes (panel b) from the galaxy centre. The y-axis show the line-of-sight velocity centred at the systemic velocity of the galaxy. The reconstructed source is shown in black, the kinematic model in red. The black dots show the derived projected velocities.}
    \label{fig:02}
\end{figure*}

The results of this lens-kinematic analysis can be visualised in three sets of figures for each lensed system. The first set shows the zeroth-, first- and second-moment maps of the data, the reconstructed source and the source kinematic model (Fig.~\ref{fig:01} and Figs.~\ref{fig:ap4} to \ref{fig:ap13}). The second set displays the data, the model, the reconstructed source and the kinematic model for some representative channel maps (Fig.~\ref{fig:03} and Figs.~\ref{fig:ap6} to \ref{fig:ap15}). In the source columns, we show the sources obtained with (gray solid contours) and without (dashed black contours) the kinematic hyperprior (see point \ref{step:3} in \ref{sec:mod_strategy}.) The third set of images shows the position-velocity (p-v) diagrams along the major and minor axis for the reconstructed sources (black contours) and their kinematic models (red contours, see Fig.~\ref{fig:02} and Figs.~\ref{fig:ap5} to \ref{fig:ap14}). The most probable a posteriori lens and kinematic parameters are listed in Tables \ref{tab:tab_lens} and \ref{tab:tab_kin}, respectively.\\
\begin{table*}
	\centering
	\caption[Lens parameters]{Mass model parameters of the gravitational lens galaxies. From left to right we list the mass density profile normalisation, position angle, axis ratio, slope, the external shear strength and its position angle.}
	\label{tab:tab_lens}
	\begin{tabular}{ccccccc} 
		\hline
		\noalign{\smallskip}
		Name & $\kappa_0$ & $\theta$ & $q$ & $\gamma$ & $\Gamma_{\mathrm{sh}}$ & $\theta_{\mathrm{sh}}$\\
			  & arcsec         & $\deg$   &  	   & 	               & &  $\deg$\\
	    \noalign{\smallskip}
		\hline
		\noalign{\smallskip}
		 SPT0113-46 &	 1.20$\pm$0.02 & 91$\pm$2 & 0.72$\pm$ 0.02 & 1.99$\pm$0.02 & 0.0013$\pm$0.0002 & 14$\pm$2\\
				      & 0.43$\pm$0.02 & 178$\pm$6 & 0.9$\pm$0.1 & 2.1$\pm$0.2\\\noalign{\vspace{1pt}}
				      & 0.15$\pm$0.02 & 37$\pm$2 & 0.8$\pm$0.1 & 2.0$\pm$0.1 \\\noalign{\vspace{1pt}}
		 SPT0345-47 & 0.36$\pm$0.05 & 128.0$\pm$0.3 & 0.71$\pm$0.04 & 1.9$\pm$0.2 & -0.027$\pm$0.004 & -6$\pm$1\\\noalign{\vspace{1pt}}
		 SPT0441-46 & 0.63$\pm$0.05 & 82$\pm$2 &0.61$\pm$0.07 & 2.0$\pm$0.2 & 0.00011$\pm$0.00001 & 37$\pm$5\\\noalign{\vspace{1pt}}
		 SPT2146-55 & 0.89$\pm$0.07 & 106$\pm$15 & 0.95$\pm$0.01 & 2.0$\pm$0.1 & 0.0136$\pm$0.0007 & -24$\pm$3\\\noalign{\vspace{1pt}}
		SPT2132-58 & 0.30$\pm$0.01 & 105$\pm$8  & 0.64$\pm$0.07 & 2.0$\pm$0.1 & -0.0610$\pm$0.003 & -11.4$\pm$0.8\\\noalign{\vspace{1pt}}
    \noalign{\smallskip}
    \hline
    \noalign{\medskip}
	\end{tabular}
\end{table*}

\begin{table*}
	\centering
	\caption[Kinematic parameters of the sources]{Kinematic parameters of the sources. The kinematic model is a rotating disc defined by the geometrical parameters ($i$ and $PA$) and the parameters defining the rotation curve, equations (\ref{eq:mult}), (\ref{eq:acrtg}), and velocity dispersion profile, equation (\ref{eq:exp}).}
	\label{tab:tab_kin}
	\begin{tabular}{ccccccccc}
		\hline
		\noalign{\smallskip}
		Name & $i$ & $PA$ & $V_{\mathrm{t}}$ & $R_{\mathrm{t}}$ & $\beta$ & $\xi$ & $\sigma_0$ & $R_{\sigma}$\\
			  & $\deg$ & $\deg$ & km s$^{-1}$ & kpc 			    &  	            & 		& km s$^{-1}$ & kpc\\
	    \noalign{\smallskip}
		\hline
		\noalign{\smallskip}
		 SPT0113-46 &	70$\pm$1 & 240$\pm$3 & 352$\pm$8 & 0.020$\pm$0.004 & 0.79$\pm$0.09 & 2.3$\pm$0.2  & 145$\pm$12 & 1.5$\pm$0.2\\\noalign{\vspace{1pt}}
		 SPT0345-47 & 53$\pm$5 & 303$\pm$6 & 200$\pm$25 & 0.68$\pm$0.09 & 1.2$\pm$0.1 & 2.8$\pm$0.3 & 123$\pm$13 & 2.2$\pm$0.3\\\noalign{\vspace{1pt}}
		 SPT0441-46 & 57$\pm$8 & 10$\pm$2 & 293$\pm$7 & 0.17$\pm$0.02 & 1.1$\pm$0.1 & 2.9$\pm$0.3 & 151$\pm$23 & 0.68$\pm$0.09 \\\noalign{\vspace{1pt}}
		 SPT2146-55 & 47$\pm$5 & 302$\pm$6 & 176$\pm$18 & 0.34$\pm$0.03 & 0.68$\pm$0.05 & 2.1$\pm$0.3 & 76$\pm$10 & 2.0$\pm$0.3\\\noalign{\vspace{1pt}}
		SPT2132-58 & 52$\pm$7 & 65$\pm$8 & 219$\pm$22 & 0.48$\pm$0.03 &.        	&	 &  50$\pm$7 & 2.6$\pm$0.3\\\noalign{\vspace{1pt}}
    \noalign{\smallskip}
    \hline
    \noalign{\medskip}
	\end{tabular}
\end{table*}

We note that four out of the five sources are lensed by a galaxy, while SPT0113-46 is lensed by a group of galaxies \citep{spilker}. When modelling SPT0113-46, we therefore also included the contribution to the lensing potential of the three closest galaxies in the form of three elliptical power-law components. We found that this was sufficient to fit the data to the noise level (see column three in Fig.~\ref{fig:03}). The lens mass parameters listed in Table \ref{tab:tab_lens} are consistent with previous analysis of the same systems \citep{spilker}.\\ 
To define the lowest contours in Figs.~\ref{fig:03} and \ref{fig:02} and the corresponding figures for the rest of the sample (Appendix \ref{appendixb}), we construct maps of the SNRs in each channel of the reconstructed source and define as reliable the pixels with SNR $\gtrsim 3$, allowing us to discriminate features in the reconstructed source from noise artefacts. The characterization of the noise in the source plane is not trivial: correlated noise features in the lens plane may be absorbed into the source reconstruction \citep{stacey20}. In particular, for the data sets used in our analysis, the uncertainties in the source surface brightness resulting from noise artefacts are, indeed, significantly larger than the uncertainties due to the lens model parameters. To estimate the uncertainties in the source plane, we, therefore, assume our maximum a posteriori source model and create mock lensed data with 100 different Gaussian realisations of the noise at the level measured in the real data. We then reconstruct the source from the mock data sets and measure the mean and standard deviation of the surface brightness in each pixel and in each spectral channel.\\

\subsection{Is a rotating disc a good model for galaxies at z $\sim$ 4.5?}
\label{sec:result:kin}
Both the figures containing the channels maps (Fig.~\ref{fig:03} and Figs.~\ref{fig:ap6} to \ref{fig:ap15}) and the p-v diagrams (Fig.~\ref{fig:02} and Figs.~\ref{fig:ap5} to \ref{fig:ap14}) show how well the rotating disc model is able to reproduce the emission of the reconstructed sources in our samples. The p-v diagrams along the major axis of the reconstructed sources (black contours) have, in fact, an S-shape, that is a typical signature of a rotating disc (see the red contours). The thickness of the p-v diagrams both along the major- and minor-axis is the result of the velocity dispersion and the instrumental resolution. At the central channel maps, close to the systemic velocity, the black contours of all galaxies have a "butterfly" shape, a typical pattern of a rotating disc.\\
We note also that on top of the bulk rotation motions two galaxies of the sample, SPT0441-46 and SPT2146-55, have anomalous kinematic components, features very common in nearby spiral galaxies \citep{fraternali01, fraternali02}. In particular, SPT0441-46 and SPT2146-55 show faint emission in the so-called forbidden regions (i.e., forbidden for rotation) of the p-v diagrams (see the black contours in the lower left quadrants in Fig.~\ref{fig:ap8} and \ref{fig:ap11}). These features are due to lag in rotation and non-circular motions and they are usually ascribed to extraplanar gas, bars or lopsided structures \citep{fraternali06, sancisi}. The inspection of the p-v diagrams and channel maps indicates that the derived kinematic parameters are robust also for these two galaxies. Due to the faint emission of the anomalous kinematic components, the kinematic models are, in fact, not influenced at all by them and are able to well reproduce the bulk of the gas in regular rotation. We also note that the presence of non-circular motions was suggested by \citet{hodge19} for a sample of DSFGs to explain the signatures of bars and non-axisymmetric structures not described by a smooth exponential brightness profile.
\\
Rotating discs have been previously observed among the DSFG population at these redshifts \citep[e.g.,][]{hodge12, carniani13, tadaki, sharda, rizzo3, fraternali20, lelli21}. However, the presence of rotation is not sufficient to rule out the merger scenario as one of the main formation channels of the DSFG population. A number of numerical simulations showed that at high-$z$ rotating gas discs can reform after gas-rich mergers \citep[e.g.,][]{robertson, hopkins13, kohandel}. We note also that mergers are not the unique processes that can explain the physical properties of DSFGs. For instance, using numerical and semi-analytical models, recent studies \citep[e.g.,][]{narayanan15, lacey, mcalpine, lagos20} showed how gas accretion, disc or bar instabilities can trigger the starburst phase and explain most properties of the DSFG population.

\subsection{Dynamical properties}
\begin{figure*}
    \centering
    \includegraphics[width=0.85\textwidth]{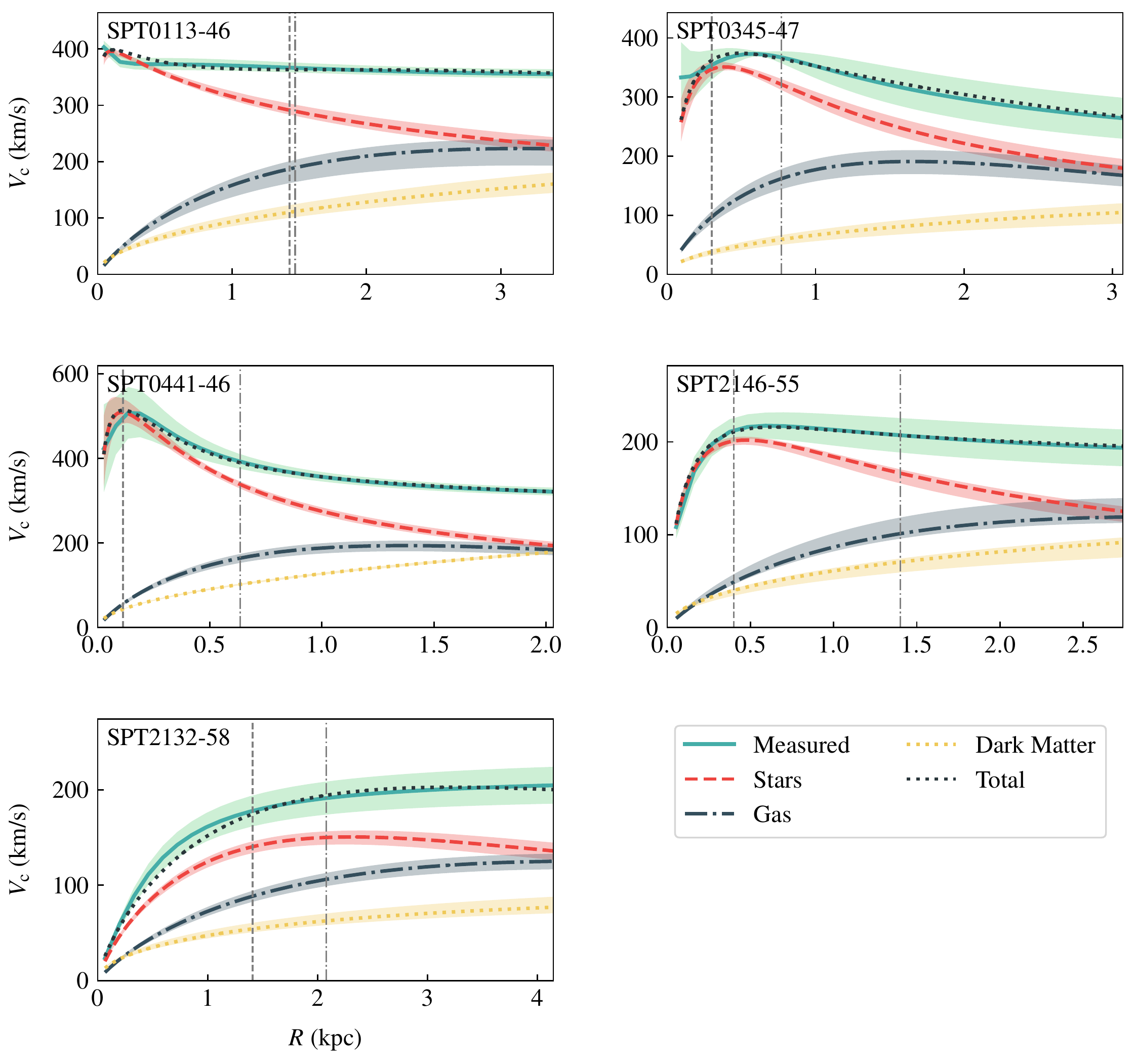}
    \caption{Rotation curve decomposition. The green solid lines show the circular velocity profiles. The black dotted lines show the best dynamical models, and the contribution from the different mass components as indicated by the legend and listed in Tables \ref{tab:dynamic1} and \ref{tab:dynamic2}. The vertical gray dashed and dotted-dashed lines show the locations of $R_{\mathrm{e}}$ (effective radius of the stellar components) and $R_{\mathrm{gas}}$ (scale radius) of the gas component.}
    \label{fig:04}
\end{figure*}

The rotation curves of the galaxies of our sample resemble those found for local spiral galaxies \citep{lelli16}. All rotation curves flatten, in fact, at large radii and show a variety of shapes in the inner regions. In particular, we can distinguish three main classes:
\begin{itemize}
    \item rotation curves that steeply rise and then decline before flattening out to the outermost measured value (SPT0113-46, SPT0441-46). These curves are typical of nearby massive spiral galaxies with stellar bulges \citep{cimatti};
    \item rotation curves with an inner slow rise and then a flattening (SPT2132-58), typical of present-day disc galaxies;
    \item rotation curves with an intermediate behaviour (SPT0345-47, SPT2146-55).
\end{itemize}
The variety of the stellar distributions in these DSFGs is, therefore, imprinted in their dynamics. In particular, we confirm a previous finding \citep{rizzo3}: some DSFGs at $z \approx 4$ show the typical signature of a bulge, that is, a bump in the inner regions of the rotation curve. \\
By performing the rotation curve decomposition described in Section \ref{sec:src_dyn}, we derive the parameters describing the three matter components, mass, gas and dark matter for each galaxy in the sample (Table \ref{tab:dynamic1}). The respective posterior distributions are shown in Fig.~\ref{fig:corner}.
Fig.~\ref{fig:04} shows for each galaxy of the sample the circular velocity and the rotation curve decomposition.\\ 

A previous estimate of the stellar mass, based on spectral energy distribution (SED) fitting, for one galaxy of our sample (SPT2146-55) is already present in the literature \citep{ma}. The reported value is $0.8^{+1.9}_{-0.6}$ 10$^{11}$ M$_{\odot}$ \citep{ma}, which, given its large uncertainties, is consistent at 1.1-$\sigma$ level with the value of 1.0$^{+0.17}_{-0.13}$ 10$^{10}$ M$_{\odot}$ found in this work (see Table \ref{tab:dynamic1}). By using the same methodology for the rotation curve decomposition, \citet{rizzo3} found a stellar mass for SPT0418-47 of 1.7$^{+0.2}_{-0.1}$ 10$^{10}$ M$_{\odot}$, consistent with the value of 9.5$\pm$3 10$^{9}$ M$_{\odot}$ derived from SED fitting \citep{debreuck19}. The agreement between the values of stellar masses derived with these two different and independent methods demonstrates the robustness of the values of $M_{\mathrm{star}}$ derived from rotation curve decomposition. 

We note that the regions probed by the [CII] emission line are baryon dominated: the ratio between the stellar and gas component to the total matter at the outermost radius ranges from 0.66$\pm$0.02 (SPT0441-46) to 0.90$\pm$0.03 (SPT0345-47). For three out of the five galaxies of our sample, global measurements of the stellar-to-halo mass ratios ($M_{\mathrm{star}}/M_{\mathrm{DM}}$) are in the range $\sim 10^{-2}$-$10^{-3}$, in agreement with theoretical expectations \citep[e.g.,][]{behroozi}. SPT0345-47 and SPT2132-58 have, instead, a $M_{\mathrm{star}}/M_{\mathrm{DM}}$ of 0.12$^{+0.2}_{-0.06}$ and 0.6$^{+0.2}_{-0.3}$, respectively. While at low-$z$, rotation curve fitting provides strong observational constraints on the stellar-to-halo mass ratios \citep{posti}, at high-$z$ constraints on the shapes of the dark matter halo profiles are challenging \citep[see discussion in][]{genzel20}. However, in the context of $\Lambda$CDM, one can rely on the predictions of cosmological simulations that consistently show that the contribution of dark matter within the inner regions of galaxies is subdominant respect to the baryonic components \citep[e.g.,][]{lovell18, teklu}. We emphasize, therefore, that our results, valid within the $\Lambda$CDM framework, show that there are small degeneracies between the baryonic and dark matter components (Fig.~\ref{fig:corner}).
\begin{table*}
	\centering
	\caption[The dynamical parameters of the sources]{Dynamical parameters of the sources. All parameters listed in columns two to six are inferred from the rotation curve decomposition assuming a Sérsic profile for the stellar component, an exponential profile for the gas disc and an NFW profile for the dark matter halo. From left to right: the stellar mass, the stellar distribution effective radius and Sérsic index, the conversion factor between the [CII] luminosities and the gas mass and the dark matter mass. Column 7 lists the best fit scale radii describing the [CII] profile. }
	\label{tab:dynamic1}
		\begin{tabular}{ccccccc}
		\hline
		\noalign{\smallskip}
		Name & $M_{\mathrm{star}}$ & $R_{\mathrm{e}}$ & $n$ & $\alpha_{\mathrm{[CII]}}$ & $M_{\mathrm{DM}}$ & $R_{\mathrm{gas}}$ \\
			  & 10$^{10}$ M$_{\odot}$ & kpc &  &   M$_{\odot}$/L$_{\odot}$ & 10$^{11}$ M$_{\odot}$ & kpc \\
		\noalign{\smallskip}
		\hline
		\noalign{\smallskip}
		 SPT0113-46 &	6$^{+2}_{-1}$ & 1.4$^{+0.6}_{-0.5}$ & 6.0$^{+0.5}_{-0.6}$ & 19$^{+4}_{-3}$ & 12$^{+6}_{-6}$ & 1.5$\pm$0.2\\\noalign{\vspace{1pt}}
		 SPT0345-47 & 2.3$^{+0.4}_{-0.3}$ & 0.30$^{+0.06}_{-0.04}$ & 1.5$^{+0.5}_{-0.3}$ & 7.3$^{+1}_{-1}$ & 2$^{+2}_{-1}$ & 0.76$\pm$0.02\\\noalign{\vspace{1pt}}
		 SPT0441-46 & 1.8$^{+0.2}_{-0.2}$ & 0.11$^{+0.02}_{-0.02}$ & 2.0$^{+0.9}_{-0.6}$ & 8$^{+1}_{-1}$ & 72$^{+8}_{-8}$ & 0.63$\pm$0.07\\\noalign{\vspace{1pt}}
		 SPT2146-55 & 1.0$^{+0.2}_{-0.1}$ & 0.39$^{+0.08}_{-0.05}$ & 1.6$^{+0.3}_{-0.2}$ & 4.6$^{+0.9}_{-0.8}$ & 1.2$^{+0.5}_{-0.5}$ & 1.4$\pm$0.1 \\\noalign{\vspace{1pt}}
		SPT2132-58 & 1.9$^{+0.3}_{-0.3}$ & 1.4$^{+0.1}_{-0.1}$ & 0.94$^{+0.05}_{-0.04}$ & 7.2$^{+0.9}_{-0.9}$ & 0.32$^{+0.23}_{-0.09}$ & 2.1$\pm$0.2\\\noalign{\vspace{1pt}}
    \noalign{\smallskip}
    \hline
    \noalign{\medskip}
	\end{tabular}
\end{table*}

In Table \ref{tab:dynamic2} we list other derived physical quantities, which are relevant to our discussion (see Section \ref{sec:discussion}), as the gas fraction $f_{\mathrm{gas}}$ and the total baryonic mass $M_{\mathrm{bar}}$. The inferred values of $M_{\mathrm{gas}}$ are in agreement, within the 2-$\sigma$ uncertainties, with the values derived from the CO luminosities, dust masses \citep{aravena}, and [CI] luminosities \citep{bothwell17}. A detailed comparison between these different gas tracers is beyond the scope of this paper and is left to a future detailed investigation.  We note that the median inferred value of $\alpha_{\mathrm{[CII]}} = 7^{+4}_{-1}$ M$_{\odot}$/L$_{\odot}$ is a factor of $\sim 4$ smaller than the value of 30 M$_{\odot}$/L$_{\odot}$ found for main-sequence galaxies by \citet{zanella}. An $\alpha_{\mathrm{[CII]}}$ of 30 M$_{\odot}$/L$_{\odot}$ is excluded by the rotation curves as the resulting $V_{\mathrm{gas}}$ (eq.~\ref{eq:vgas}) would be larger than the observed $V_{\mathrm{c}}$. In addition, by assuming $\alpha_{\mathrm{[CII]}} = 30$ M$_{\odot}$/L$_{\odot}$, the resulting values of $M_{\mathrm{gas}}$ would systematically overestimate the previous measurements from different gas tracers \citep{aravena, bothwell17}. Values of $\alpha_{\mathrm{[CII]}}$ of 10 M$_{\odot}$/L$_{\odot}$ were previously assumed for DSFGs in the literature \citep[e.g.,][]{swinbank122, gullberg}, based on the comparison between CO(1-0) and [CII] luminosities \citep[e.g.,][]{stacey, debreuck11}. Also, using theoretical prescriptions, \citet{sommovigo} found that $\alpha_{\mathrm{[CII]}}$ depends on the star-formation mode and it is typically $\lesssim 10$ M$_{\odot}$/L$_{\odot}$ in star-forming galaxies at $z \gtrsim 4$.

\begin{table*}
	\centering
	\caption[{Derived physical properties}]{Derived physical properties of the sources. From left to right the gas mass, the fraction of total baryonic mass in gas, the total baryonic mass, the baryonic effective radius, the gas depletion time and the disc-scale height.}
	\label{tab:dynamic2}
	\begin{tabular}{ccccccc} 
		\hline
		\noalign{\smallskip}
		Name & $M_{\mathrm{gas}}$ & $f_{\mathrm{gas}}$ & $M_{\mathrm{bar}}$ & $R_{\mathrm{bar}}$ & $t_\mathrm{dep}$ & $h$\\
			  & 10$^{10}$ M$_{\odot}$ &  & 10$^{10}$ M$_{\odot}$ & kpc & Myr & pc\\
	    \noalign{\smallskip}
		\hline
		\noalign{\smallskip}
		 SPT0113-46 &	4.3$^{+1.0}_{-0.9}$ & 0.39$^{+0.11}_{-0.09}$ & 11$^{+1}_{-1}$ & 2.2$^{+0.4}_{-0.2}$ & 357$\pm$ 73 & 100$^{+86}_{-36}$\\\noalign{\vspace{1pt}}
		 SPT0345-47 & 1.7$^{+0.4}_{-0.4}$ & 0.42$^{+0.08}_{-0.09}$ & 4.0$^{+0.3}_{-0.3}$ & 0.64$^{+0.07}_{-0.05}$ & 12$\pm$ 2 & 351$^{+88}_{-88}$\\\noalign{\vspace{1pt}}
		 SPT0441-46 & 1.4$^{+0.2}_{-0.2}$ & 0.45$^{+0.05}_{-0.06}$ & 3.2$^{+0.2}_{-0.2}$ & 0.35$^{+0.06}_{-0.05}$ & 22$\pm$4 & 225$^{+105}_{-84}$\\\noalign{\vspace{1pt}}
		 SPT2146-55 & 1.2$^{+0.3}_{-0.2}$ & 0.54$^{+0.07}_{-0.08}$ & 2.2$^{+0.2}_{-0.2}$ & 1.3$^{+0.2}_{-0.2}$ & 17$\pm$4 & 324$^{+99}_{-73}$ \\\noalign{\vspace{1pt}}
		SPT2132-58 & 1.9$^{+0.3}_{-0.3}$ & 0.50$^{+0.07}_{-0.07}$ & 3.9$^{+0.4}_{-0.3}$ & 2.5$^{+0.1}_{-0.1}$ & 20$\pm$4 & 368$^{+90}_{-74}$ \\\noalign{\vspace{1pt}}
    \noalign{\smallskip}
    \hline
    \noalign{\medskip}
	\end{tabular}
\end{table*}

\subsection{Source continuum properties and SFR} \label{sec:continuum}
In order to derive the star formation properties of the sources we model their continuum emission in a narrow spectral range close to the redshifted [CII] emission and use the strong lensing magnification factor (Table \ref{tab:sfr}) to compute the intrinsic infrared luminosity from the observed values taken from the literature \citep[][column two of Table \ref{tab:sfr}]{aravena}. Specifically, these infrared luminosities were obtained from SED fitting of far-infrared and sub-millimeter observations, covering the range 250 to 3000 $\mu$m \citep{aravena}. 
By applying the Kroupa Initial-Mass-Function (IMF) conversion factor of $1.48 \times 10^{-10}$~M$_{\odot}$~yr$^{-1}$~L$_{\odot}^{-1}$\citep{kennicutt}, we then derived the SFR for each source (column four of Table \ref{tab:sfr}). 

\begin{table}
	\centering
	\caption[SFR properties]{SFR and [CII] luminosities of the sources. Column two: the observed infrared luminosity from \citet{aravena}. Column three: the magnification factor of the continuum in the infrared bands. Column four: star-formation rate derived for a Kroupa IMF. Column five: intrinsic [CII] luminosities.}
	\label{tab:sfr}
	\begin{tabular}{cccccccc} 
		\hline
		\noalign{\smallskip}
		Name & $L_{\mathrm{IR,obs}}$ & $\mu$ & SFR & $L_{\mathrm{[CII]}}$ \\
			  & 10$^{13}$ L$_{\odot}$ & &  10$^{2}$M$_{\odot}$ yr$^{-1}$ & 10$^{9}$ L$_{\odot}$\\
	    \noalign{\smallskip}
		\hline
		\noalign{\smallskip}
		 SPT0113-46 & 3.0$\pm$0.5 & 36$\pm$4 & 1.2$\pm$0.3 & 2.3$\pm$0.1\\\noalign{\vspace{1pt}}
		 SPT0345-47 & 13$\pm$2 & 14$\pm$1 & 13$\pm$3 & 2.3$\pm$0.2 \\\noalign{\vspace{1pt}}
		 SPT0441-46 & 4.8$\pm$0.9 & 10.8$\pm$0.5 & 7$\pm$1 & 1.8$\pm$0.1 \\\noalign{\vspace{1pt}}
		 SPT2146-55 & 3.6$\pm$0.8  & 7.8$\pm$0.2 & 7$\pm$1 & 2.6$\pm$0.2\\\noalign{\vspace{1pt}}
		SPT2132-58 & 4.2$\pm$0.7 & 6.3$\pm$0.4 & 10$\pm$2 & 3.7$\pm$0.7\\\noalign{\vspace{1pt}}
    \noalign{\smallskip}
    \hline
    \noalign{\medskip}
	\end{tabular}
\end{table}

\subsubsection{Position in the SFR - $M_{\mathrm{star}}$ plane}
\label{sec:srf_mass}

For most star-forming galaxies, there is a tight correlation between their SFR and their stellar mass \citep{brinchmann}, the so-called main sequence. Several studies \citep[e.g.,][]{noeske, whitaker, steinhardt, tasca} showed that the main sequence holds from $z = 0$ out to $z \sim 6$, with a redshift evolution of its normalization. Also, there are starbursts galaxies, characterized by significantly higher SFR than normal main-sequence galaxies. While starburst galaxies are rare in the local Universe \citep{renzini}, they constitute a
moderate percentage \citep[$\approx$ 15 per cent,][]{bisigello, caputi} of all galaxies at $z\gtrsim 2$, contributing up to 50 per cent of the cosmic SFR density at $z\sim 4$ \citep{caputi}.\\
Starting from the physical properties derived in the previous section, we now explore the location of the source galaxies in the SFR - $M_{\mathrm{star}}$ diagram. We find that four out of five galaxies in our sample are consistent with the starburst sequence, while SPT0113-46 is a main-sequence galaxy. These findings are not surprising, given that the selection criteria used for the identification of the SPT DSFGs are based on the observed flux rather than the intrinsic one \citep{vieira10, vieira}. SPT0113-46 has, indeed, an observed infrared luminosity of $L_{\mathrm{IR,obs}}$ which is similar to the rest of the sample (see column two in Table \ref{tab:sfr}). However, the large magnification factor of $\sim 36$ leads to an intrinsic luminosity $L_{\mathrm{IR}}$ and SFR which are on average a factor of five below the other galaxies in the sample.

\begin{figure}
    \centering
    \includegraphics[width=\columnwidth]{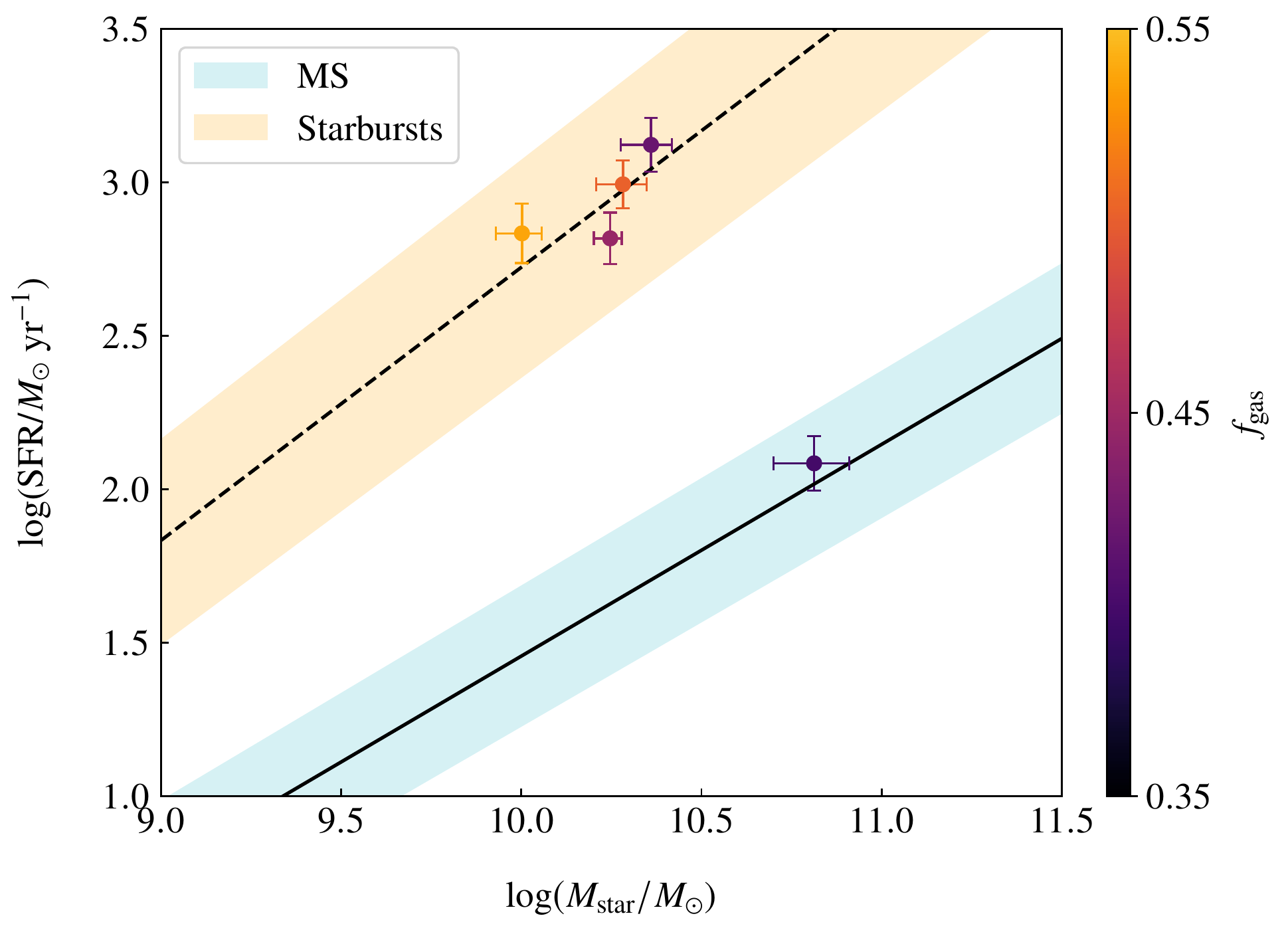}
    \caption[SFR - $M_{\mathrm{star}}$ plane]{
    Location on the SFR - $M_{\mathrm{star}}$
    of the source galaxies in our sample (circles), with markers colour-coded according to their gas fraction. The solid black line and the blue area show the best-fit and the 1-$\sigma$ scatter for main-sequence galaxies at $z\sim$ 4 - 5 from \citet{caputi}, respectively. The dotted black line and the orange area show the starburst sequence.}
    \label{fig:ms}
\end{figure}

\section{Discussion}
\label{sec:discussion}
In this section, we investigate the dynamical and structural properties of the reconstructed sources. For the rest of this paper, we also include, if not otherwise stated, four non-lensed and one lensed DSFGs in our sample (see Table \ref{tab:sample_ext}). These sources are the only $4 \lesssim z \lesssim 5$ DSFGs with beam-smearing corrected CO \citep{sharda} or [CII] kinematic measurements \citep{fraternali20, rizzo3, lelli21} and without clear merging signatures.

\begin{table*}
	\centering
	\caption{Properties of the galaxies added to our sample. The kinematic properties in columns three to five are taken from the references shown in column six. The SFR value (converted to a Kroupa IMF) in column seven are taken from the references in column eight.}
	\label{tab:sample_ext}
	\begin{tabular}{cccccccc} 
		\hline
		\noalign{\smallskip}
		Name & $z$ & $V_{\mathrm{flat}}$ & $\sigma_{\mathrm{ext}}$ & $\sigma_{\mathrm{m}}$ & Reference & SFR & Reference \\
			  &  & km s$^{-1}$ & km s$^{-1}$ & km s$^{-1}$ & & M$_{\odot}$ yr$^{-1}$\\
	    \noalign{\smallskip}
		\hline
		\noalign{\smallskip}
		 SPT0418-47 & 4.23 & 259$\pm1$ & 18$\pm1$ & 32$\pm1$ & \citet{rizzo3} & 352$\pm65$ & \citet{rizzo3}\\\noalign{\vspace{1pt}}
		 AzTEC-1 & 4.34 & 200$\pm27$ & 30$\pm8$ & 40$^{+20}_{-10}$ & \citet{sharda} & 1261$^{+38}_{-309}$ & \citet{tadaki}\\\noalign{\vspace{1pt}}
		 J1000+0234 & 4.54 & 538$\pm40$ & $\lesssim 60$ & $\lesssim 60$ & \citet{fraternali20} & 468$^{+1276}_{-340}$ & \citet{gomez}\\\noalign{\vspace{1pt}}
		  AzTEC/C159 & 4.57 & 506$^{+151}_{-76}$ & 16$\pm13$ & 16$\pm13$ & \citet{fraternali20} & 787$^{+223}_{-180}$ & \citet{gomez} \\\noalign{\vspace{1pt}}
		  ALESS073.1 & 4.75 & 309$^{+47}_{-66}$ & 32$^{+10}_{-9}$ & 47$^{+15}_{-28}$ & \citet{lelli21} & 579$\pm$116 & \citet{circosta}\\\noalign{\vspace{1pt}}
    \noalign{\smallskip}
    \hline
    \noalign{\medskip}
	\end{tabular}
\end{table*}

\subsection{$V/\sigma$ ratio for non-merging DSFGs}
\label{sec:bias}
\begin{figure*}
    \centering
    \includegraphics[width=0.5\textwidth]{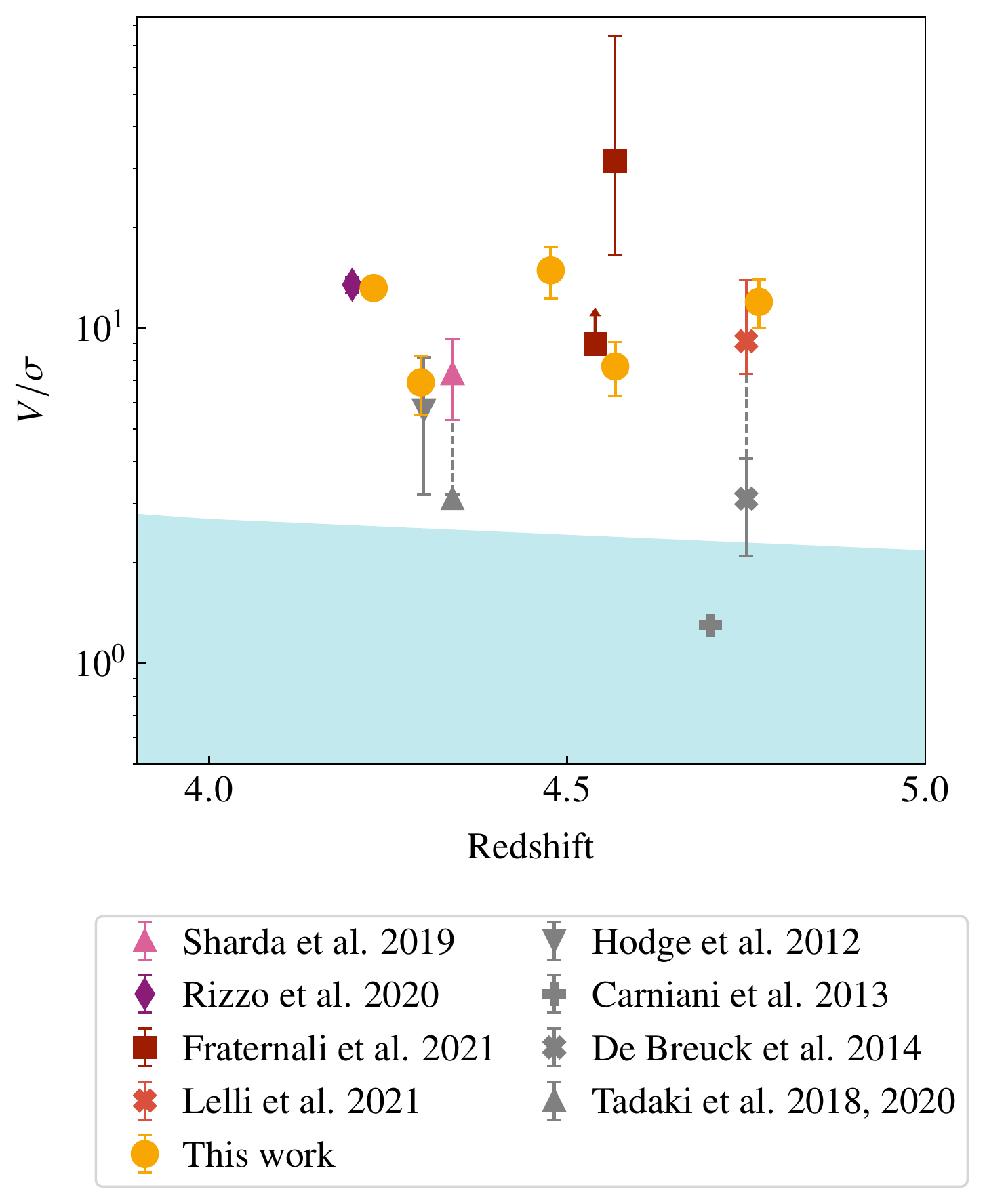}
    \caption[Rotational support]{$V/\sigma$ versus redshift. The $V/\sigma$ for our sample (yellow circles), defined as $V_{\mathrm{flat}}/\sigma_{\mathrm{ext}}$ and for the non-merging DSFGs from literature (coloured markers, Table \ref{tab:sample_ext}), as indicated in the legend. Note that the redshift of SPT0418-47 (orange diamond) is shifted by -0.02 for a better visualisation of all the points. The light-blue area shows the region covered by theoretical studies \citep{pillepich, dekel, zolotov15, hay}. The gray markers show the $V/\sigma$ measurements that may be affected by systematic bias (see Section \ref{sec:bias}). The dashed lines join the markers for which two different $V/\sigma$ estimates for the same galaxy are available.}
    \label{fig:vsigma}
\end{figure*}

Recent observational and theoretical studies \citep[e.g.,][]{swinbank17, wisnioski, ubler19, turner, zolotov15, pillepich, dekel20} have suggested that many high-$z$ galaxies, despite being rotationally supported systems, have intrinsic velocity dispersions which are much higher than those of local galaxies. The level of rotational support relative to the amount of turbulence in a galaxy is generally quantified with the ratio between the rotation velocity and the velocity dispersion, $V/\sigma$. We compute the $V/\sigma$ ratios for our sample using two definitions: $V_{\mathrm{max}}/\sigma_{\mathrm{m}}$, the ratio between the maximum rotation velocity and the median velocity dispersion and $V_{\mathrm{flat}}/\sigma_{\mathrm{ext}}$, the ratio between the flat part of the rotation velocity and the velocity dispersion at outer radii (Table \ref{tab:globalkin}). \\
In Fig.~\ref{fig:vsigma}, the $V_{\mathrm{flat}}/\sigma_{\mathrm{ext}}$ ratios are plotted as a function of redshift and compared with theoretical predictions \citep{pillepich, dekel, zolotov15, hay}. We find that the ten galaxies of our extended sample have $V/\sigma$ ratios which are sistematically larger than any current theoretical prediction. In particular, the median $V/\sigma$ ratio of the sample is 11$^{+4}_{-3}$, which is a factor of 4 larger than the highest value of $\approx 3$ predicted by TNG50 \citep{pillepich}. This result confirms the findings presented in \citet{rizzo3}, \citet{fraternali20} and \citet{lelli21} that DSFGs at these high redshifts may be significantly less turbulent than expected. However, we note that the velocity dispersions measured for the galaxies in our sample are typically higher than the thermal velocity dispersions of $\approx 10$ km s$^{-1}$ expected for a gas at temperatures $T \lesssim 10^4$ K \footnote{The gas velocity dispersions, $\sigma$ is the sum in quadrature of two contributions, $\sigma^2 = \sigma_\mathrm{th}^2 + \sigma_\mathrm{turb}^2$, where $\sigma_\mathrm{turb}$ is the velocity dispersion due to the turbulence and $\sigma_\mathrm{th}$ is the velocity dispersions due to the thermal motions of the particles within the gas \citep{cimatti}. In particular, $\sigma_\mathrm{th}$ depends on the temperature $T$ of the fluid: $\sigma_\mathrm{th} = \sqrt{k_{\mathrm{B}}T/(\widetilde{m} m_\mathrm{p})}$, where $k_{\mathrm{B}}$ is the Boltzmann constant, $\widetilde{m}$ is the mean average mass of the particle in the fluid in units of the proton mass, while $m_\mathrm{p}$ is the mass of the proton.}.
These results imply a significant level of turbulence in the ISM of these galaxies, which is most plausibly \citep[e.g.][]{ubler19, krum, varidel} related to either star-formation feedback (e.g., momentum injection by supernova explosions, stellar winds and expansion of HII regions) or gravitational phenomena (e.g., gas accretion, galaxy interactions and gravitational instabilities). Both type of mechanisms may play an important role in driving turbulent motions also in nearby galaxies \citep[see the discussion in][]{arribas, bacchini, varidel}. Due to the high level of star-formation and the significant gas fraction, it is expected that both feedback and gravity-driven turbulence are more significant for high-$z$ galaxies \citep{hung, pillepich}.
In addition to the extended sample of DSFGs shown in Table \ref{tab:globalkin}, kinematic studies for two additional DSFGs can be found in the literature \citep{hodge12, carniani13}. For completeness, in Fig.\ref{fig:vsigma}, we plot the $V/\sigma$ ratio including these objects. However, due to the potential bias affecting these measurements (see below), all quantitative consideration and discussion in this Section, are done including only kinematic measurements derived from 3D analysis.\\
The derivation of the intrinsic values of $V$ and $\sigma$ requires to properly account for the beam using forward modelling techniques that fit the data in the native 3D space. However, some high-$z$ dynamical analysis were performed on the projected velocity and velocity dispersion maps \citep[2D analysis, e.g.,][]{carniani13, debreuck14}. In this case, the velocity dispersions were beam-smearing corrected a posteriori by employing empirical relations \citep[e.g.,][]{swinbank12}. \citet{lelli21} showed that the velocity dispersions obtained from an a-posteriori beam-smearing correction of the 2D maps are still biased to larger values: from a 2D study of the galaxy ALESS073.1, \citet{debreuck14} have inferred a value of $\sigma$ which at outer regions is $\sim 2$ times larger than the one derived by \citet{lelli21} from a 3D analysis of the same galaxy\footnote{In ALESS073.1, the velocity dispersions range from 10 to 40 km s$^{-1}$ in \citet{lelli21} and from 40 to 50 km s$^{-1}$ in \citet{debreuck14} at $R \gtrsim 2$ kpc, with average values of 22 km s$^{-1}$ and 40 km s$^{-1}$, respectively.}. \\
Another source of potential bias, is related to assumptions made on the velocity dispersion. For example, \citet{tadaki, tadaki20} analyzed the CO(4-3) and [CII] kinematics of AzTEC-1 using a 3D technique \citep{bouche15}. In both cases, the authors assumed a constant velocity dispersion across the disk, finding a value of $\sigma \sim$ 74 km s$^{-1}$ from the CO(4-3) and [CII] data. Relaxing this assumption, \citet{sharda} found, instead, a strong decrease of the velocity dispersion with increasing galactocentric radius (from $\sim 80$ km s$^{-1}$ in the inner regions to $\sim 30$ km s$^{-1}$ at outer radii) by performing a 3D analysis of the same CO(4-3) data employed in \citet{tadaki}. The analysis in \citet{sharda} demonstrated, therefore, that by fixing a constant velocity dispersion, one infers a value of $\sigma$ which is mainly informed by the central regions of the galaxy. For consistency with our analysis (see Section \ref{sec:src_kin}), only the measurements from \citet{sharda} are included in the extended sample and the discussion in the following sections.

\subsection{Drivers of turbulence in $z \sim 4$ non-merging DSFGs}
\label{sec:turbulence}
\begin{figure*}
    \centering
    \includegraphics[width=\textwidth]{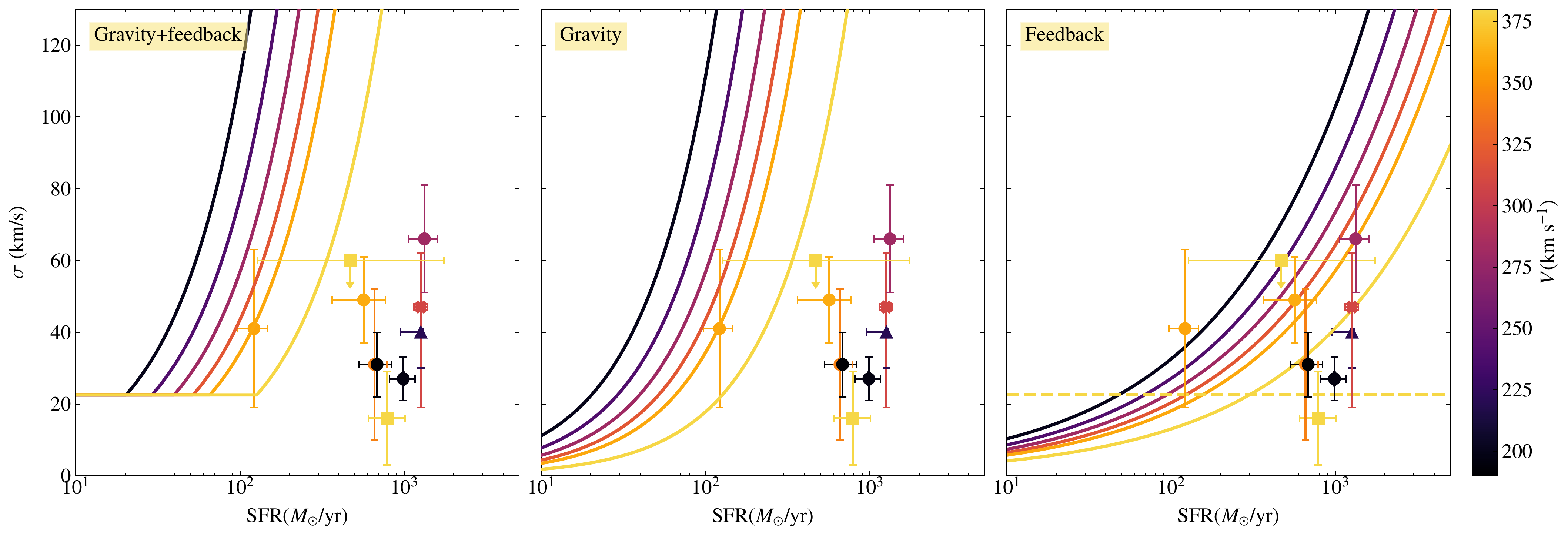}
    \caption{The markers show the $\sigma_{\mathrm{m}}$ (column three in Table \ref{tab:globalkin}) and SFR (column four in Table \ref{tab:sfr}) for our sample (circles) and for the galaxies from \citet[][diamond]{rizzo3}, \citet[][triangle]{sharda}, \citet[][squares]{fraternali20} and \citet[][cross]{lelli21}. All markers are colour-coded according to $V_{\mathrm{flat}}$ (column five in Table \ref{tab:globalkin} and column three in Table \ref{tab:sample_ext}). In the four panels, the curves show the predictions of the relation between velocity dispersions and SFR in the analytic model developed by \citet{krum}, obtained with the parameters valid for high-$z$ galaxies. The left and medium panels show the models that include gravitational instabilities plus stellar feedback and only gravitational mechanisms as the main drivers of turbulence within galaxies, respectively. The right panel shows models with only stellar feedback-driven turbulence: "Feedback, fixed $Q$" (solid lines) and "Feedback, fixed $\epsilon_{\mathrm{ff}}$" (dashed lines).}
    \label{fig:sfr_sigma}
\end{figure*}

We now compare our observations with an analytical model by \citet{krum} that takes into account gravitational mechanisms, stellar feedback or both to describe the evolution of some observed properties of star-forming galaxies. For their models, \citet{krum} assumed a rotating disc, made up of gas, stars and a spheroidal dark matter halo, in vertical hydrostatic and energy equilibrium. The dissipation of turbulence is counteracted by the injection of energy due to supernova explosions and the release of gravitational potential energy due to the inward flow of gas driven by non axisymmetric torques and gas accretion. The inclusion of either the two turbulence-driving mechanisms or only one of the two results in different relations between the velocity dispersions and the SFR. In the three panels in Fig. \ref{fig:sfr_sigma}, we plot the $\sigma$-SFR ralations obtained by assuming the fiducial parameters\footnote{A parameter defining the models by \citet{krum} is the effective gas fraction at the mid-plane, $f_{\mathrm{gas, P}} = 1.5 f_{\mathrm{gas}}$ \citep{ubler19}. In Fig. \ref{fig:sfr_sigma}, we show the $\sigma$ - SFR relations obtained by assuming the fiducial value $f_{\mathrm{gas, P}} = 0.7$. However, we note that our results do not change by assuming values of $f_{\mathrm{gas, P}}$ in the full range (0.6 - 0.8) probed by our sample.} for the high-$z$ galaxies shown in Table 3 of \citet{krum}, but with six different values of circular velocities matching those measured in our extended sample. The parameters defining the behaviour of $\sigma$ in these models are the Toomre parameter $Q$\footnote{The Toomre parameter considered in \citet{krum} is defined as $Q = \frac{\Sigma_{\mathrm{g}}}{\Sigma_{\mathrm{g}} +[2\sigma^2/(\sigma^2+\sigma_{\mathrm{star}}^2)] \Sigma_{\mathrm{star}}}\frac{\kappa \sigma}{\pi G\Sigma_{\mathrm{g}}}$, where $\Sigma_{\mathrm{g}}$ and $\Sigma_{\mathrm{star}}$ are the gas and stellar surface densities, $\sigma_{\mathrm{star}}$ is the the stellar velocity dispersion and $\kappa$ is the epicyclic frequency \citep{romeo}.} and the star formation efficiency per free-fall time $\epsilon_{\mathrm{ff}}$:
\begin{itemize}
    \item in the "Gravity + feedback" model (left panel in Fig. \ref{fig:sfr_sigma}), the turbulence is driven by both stellar feedback and gravitational instabilities. The former are able to sustain the turbulence when $Q > 1$ (flat part of the curves at $\sigma \lesssim 25$ km s$^{-1}$), while the latter play a key role for $Q = 1$. In this model, $\epsilon_{\mathrm{ff}}$ is kept constant at 0.015. We note that the Toomre parameter for our sample has a median value of 1.1, going from 0.7$^{+0.4}_{-0.2}$ (SPT0441-46) to 1.5$^{+0.2}_{-0.4}$ (SPT2146-55).
    \item The "Gravity" model (medium panel in Fig. \ref{fig:sfr_sigma}) includes only the gravitational instabilities as drivers of turbulence within the model galaxies. The assumptions are similar to the "Gravity + feedback" model, except that Q is always equal to 1.
    \item The "Feedback, fixed $\epsilon_{\mathrm{ff}}$" (dashed lines in the right panel in Fig. \ref{fig:sfr_sigma}) model is similar to the so-called ‘self-regulated-system' model \citep{ostriker}. In this case, the star formation is part of a self-regulating cycle where the momentum injected to the ISM by star formation balances the gravitational force confining the ISM gas in the disc, without any injection of gravitational potential energy. $\epsilon_{\mathrm{ff}}$ is kept constant at 0.015 and $Q$ is left free to vary.
    \item Also in the "Feedback, fixed $Q$" model (solid lines in the right panel in Fig. \ref{fig:sfr_sigma}), stellar feedback is the only driving mechanisms of turbulence. Similarly to the model developed by \citet{faucher},  the star-formation efficiency $\epsilon_{\mathrm{ff}}$ is not constant and vary as a function of different properties of the galaxy, while $Q = 1$.
\end{itemize}
The comparison between the values of $\sigma$ and SFR for the galaxies shown in Fig.~\ref{fig:sfr_sigma} and these analytical models indicate that stellar feedback is able to sustain the measured turbulence (see the right panel in Fig. \ref{fig:sfr_sigma}). The two models (left and medium panels) including gravitationally-driven mechanisms overestimate the velocity dispersions by factors from $\approx$ 7 (SPT0418-47) to $\approx$ 40 (SPT2132-58) for seven out of ten galaxies at the corresponding values of SFR. The only exceptions are SPT0113-46 that has a position consistent with all four models shown in  Fig. \ref{fig:sfr_sigma} and J1000+0234 for which only an upper limit for the measured velocity dispersion is available \citep{fraternali20}. This result is at odds with recent works \citep{krum, johnson, ubler19} finding that gravitational instabilities are fundamental to explain the position of their galaxies in the $\sigma$-SFR plane with respect to the models of \citet{krum}. The reason for this discrepancy may be ascribed to the different galaxy populations and redshifts of the observed galaxies: starbursts at $z \sim 4.5$ in this paper, normal star-forming galaxies at $1 \lesssim z \lesssim 2$ in \citet{johnson} and \citet{ubler19} (see also discussion in Section \ref{sec:sigmaz}). The major role of stellar feedback over other mechanisms in driving the observed properties of DSFGs at $z \sim$ 4 - 5 was recently emphasized by \citet{spilker20}.

\begin{table*}
	\centering
	\caption[Global kinematic parameters]{Global kinematic parameters of the sources. Column two: the maximum rotation velocity. Column three: the median velocity dispersion. Column four: the ratio between $V_{\mathrm{max}}$ and $\sigma_{\mathrm{m}}$. Column five: the rotation velocity in the flat part of the rotation curve. Column six: the velocity dispersion in the external regions ($R \gtrsim R_{\mathrm{e}}$). Column seven: the ratio between $V_{\mathrm{flat}}$ and $\sigma_{\mathrm{ext}}$.}
	\label{tab:globalkin}
	\begin{tabular}{ccccccc} 
		\hline
		\noalign{\smallskip}
		Name & $V_{\mathrm{max}}$ & $\sigma_{\mathrm{m}}$ & $V_{\mathrm{max}}/\sigma_{\mathrm{m}}$ & $V_{\mathrm{flat}}$ & $\sigma_{\mathrm{ext}}$ & $V_{\mathrm{flat}}/\sigma_{\mathrm{ext}}$\\
	    \noalign{\smallskip}
		\hline
		\noalign{\smallskip}
		 SPT0113-46 & 382$\pm$9 & 41$\pm$22 & 9.2$\pm$4.8 & 358$\pm$3  & 27$\pm$1 & 13.2$\pm$0.6\\\noalign{\vspace{1pt}}
		 SPT0345-47 & 373$\pm$5 & 66$\pm$15 & 5.6$\pm$1.2 & 280$\pm$25 & 40$\pm$7 & 6.9$\pm$1.4 \\\noalign{\vspace{1pt}}
		 SPT0441-46 & 489$\pm$67 & 31$\pm$21 & 15.8$\pm$10.8 & 342$\pm$4 & 23$\pm$4 & 14.9$\pm$2.6 \\\noalign{\vspace{1pt}}
		 SPT2146-55 & 217$\pm$13 & 31$\pm$11 & 7.0$\pm$2.6 & 194$\pm$7 & 20$\pm$2 & 9.8$\pm$1.2\\\noalign{\vspace{1pt}}
		SPT2132-58 & 199$\pm$18 & 27$\pm$6 & 7.2$\pm$1.7 & 196$\pm$14 & 16$\pm$3 &  12$\pm$2 \\\noalign{\vspace{1pt}}
    \noalign{\smallskip}
    \hline
    \noalign{\medskip}
	\end{tabular}
\end{table*}

\subsubsection{Velocity dispersions and star-formation efficiencies}
\begin{figure}
    \centering
    \includegraphics[width=\columnwidth]{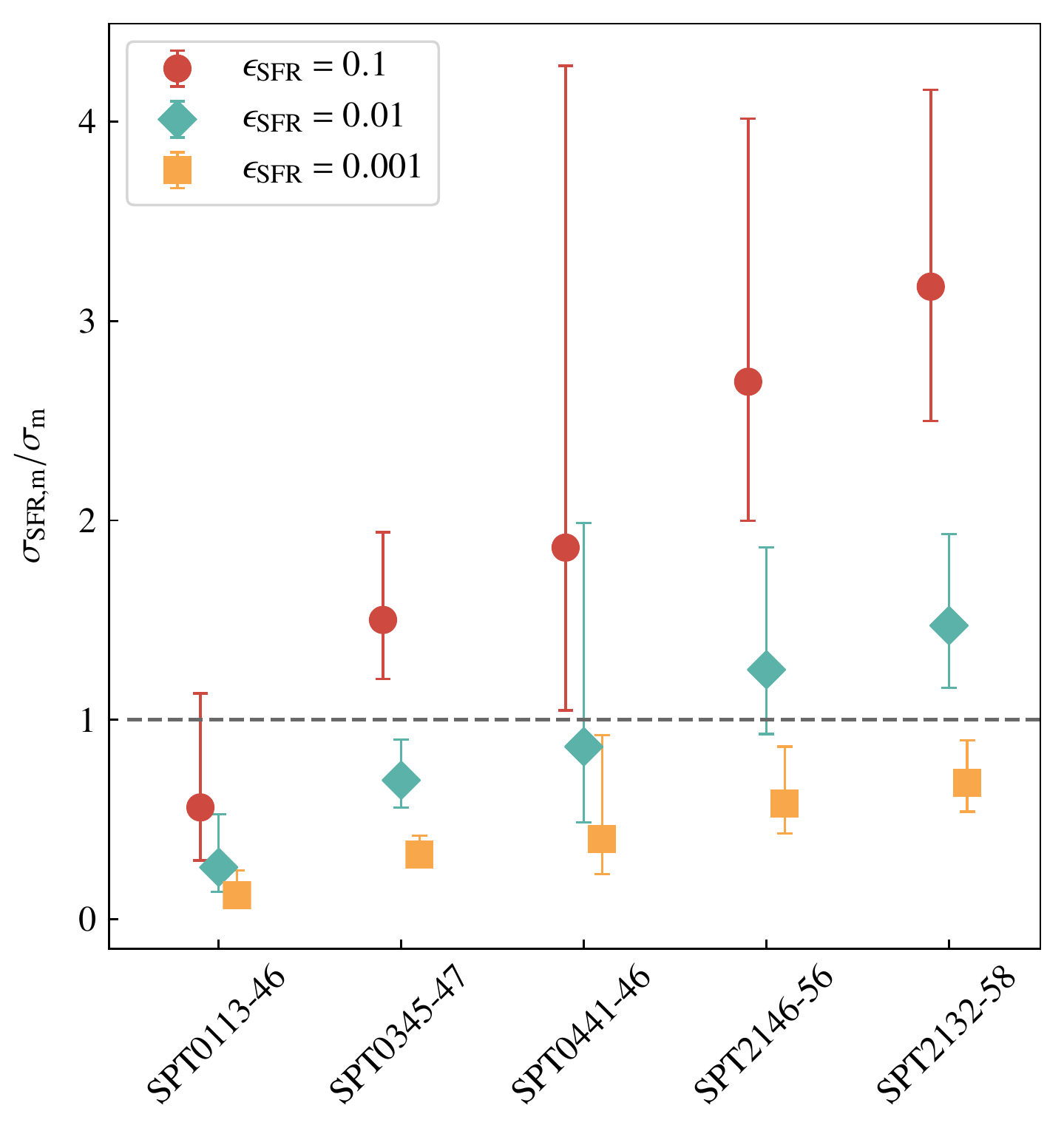}
    \caption{Ratio between the velocity dispersion expected from energy injection by the stellar feedback, equation (\ref{eq:sigma_sfr}), and the measured velocity dispersion $\sigma_{\mathrm{m}}$ (column three in Table \ref{tab:globalkin}). The three markers for each galaxy show the ratios obtained with three different values of $\epsilon_{\mathrm{SN}}$, as indicated in the legend.}
    \label{fig:ratio}
\end{figure}

As a further test, in this section we estimate whether the velocity dispersions measured for our sample can be explained by the energy injected by supernova explosions using simple and reasonable assumptions.
Following \citet{rizzo3} and \citet{fraternali20}, the velocity dispersion due to the transferring of the supernova energy to the ISM is
\begin{equation}
\sigma_{\mathrm{SFR}} = 58 \left(
\frac{\epsilon_{\mathrm{SN}}}{0.1} 
\frac{\mathrm{SFR}}{300 \mathrm{M_{\odot} yr^{-1}}}
\frac{h}{200 \mathrm{pc}}
\right)^{1/3}
\left(
\frac{M_\mathrm{gas}}{10^{10} \mathrm{M_{\odot}}}
\right)^{-1/3} \mathrm{km s^{-1}},
\label{eq:sigma_sfr}
\end{equation}
where $\epsilon_{\mathrm{SN}}$ is the efficiency of transferring kinetic energy from supernova feedback to the ISM and $h$ is the disc scale height. Equation (\ref{eq:sigma_sfr}) is obtained by assuming a supernova rate of 0.01 $M_{\odot}^{-1}$, valid for a Kroupa IMF \citep{tamburro}. Since the calculation of the disc scale height is not trivial \citep[see discussion in][]{bacchini19}, we use an analytical approximation and leave the exact estimate to a future work. The scale height of the vertical distribution of a gas disc in hydrostatic equilibrium can be approximated \citep{bacchini19} as
\begin{equation}
    h(R) = \frac{\sigma(R)}{\sqrt{4 \pi G [\rho(R)+\rho_{\mathrm{rot}}(R)]}}
    \label{eq:h}
\end{equation}
with 
\begin{equation}
    \rho_{\mathrm{rot}}(R) = -\frac{1}{2 \pi G} \frac{V_{\mathrm{c}(R)}}{R} \frac{\partial{V_{\mathrm{c}}(R)}}{\partial{R}}
    \label{eq:rorot}
\end{equation}
Since in this approximation, the self-gravity of the gas is not included, $\rho(R)$ is the density profile of the stellar and dark matter component and $\rho_{\mathrm{rot}}(R)$ is obtained by considering equations (\ref{eq:vstar}) and (\ref{eq:vdm}).\\
The values of the median $h$ for our sample are listed in column seven of Table \ref{tab:dynamic2}. In Fig.~\ref{fig:ratio}, we show the ratios between the expected values of $\sigma_{\mathrm{SFR}}$ and the measured $\sigma_{\mathrm{m}}$ (column three in Table \ref{tab:globalkin}). The $\sigma_{\mathrm{SFR}}$ values are calculated by using three values of $\epsilon_{\mathrm{SN}}$ equal to 0.001, 0.01 and 0.1, typical of observed nearby \citep{bacchini} and simulated galaxies \citep[e.g.,][]{martizzi16, ohlin}. For all galaxies of our sample the ratios between $\sigma_{\mathrm{m}}$ and one of the three values of $\sigma_{\mathrm{SFR}}$ are $\approx 1$, confirming that the turbulent motions can be easily driven by supernova explosions (see Fig. \ref{fig:ratio}). Larger values of $\epsilon_{\mathrm{SN}}$, 0.8-1, are, instead, considered an indication that other physical mechanisms, in addition to stellar feedback, drive the observed turbulent motions \citep{utomo, tamburro}. We note that three galaxies in the sample (SPT0113-46, SPT0345-47, SPT0441-46) have values of velocity dispersions fully consistent with $\sigma_{\mathrm{SFR}}$ at $\epsilon_{\mathrm{SN}} = 0.1$  within 1.5-$\sigma$. For two galaxies (SPT2146-55, SPT2132-58), the $\sigma_{\mathrm{SFR}}$ values at $\epsilon_{\mathrm{SN}} = 0.1$ overestimate $\sigma_{\mathrm{m}}$ by a factor of $\approx$ 3, indicating that low efficiency values $0.01 \lesssim \epsilon_{\mathrm{SN}} \lesssim 0.001$ are needed, similarly to what found both for high- and low-$z$ star-forming galaxies \citep{fraternali20, bacchini}.

\subsection{Turbulence across cosmic time}
\label{sec:sigmaz}
\begin{figure*}
    \centering
    \includegraphics[width=0.6\textwidth]{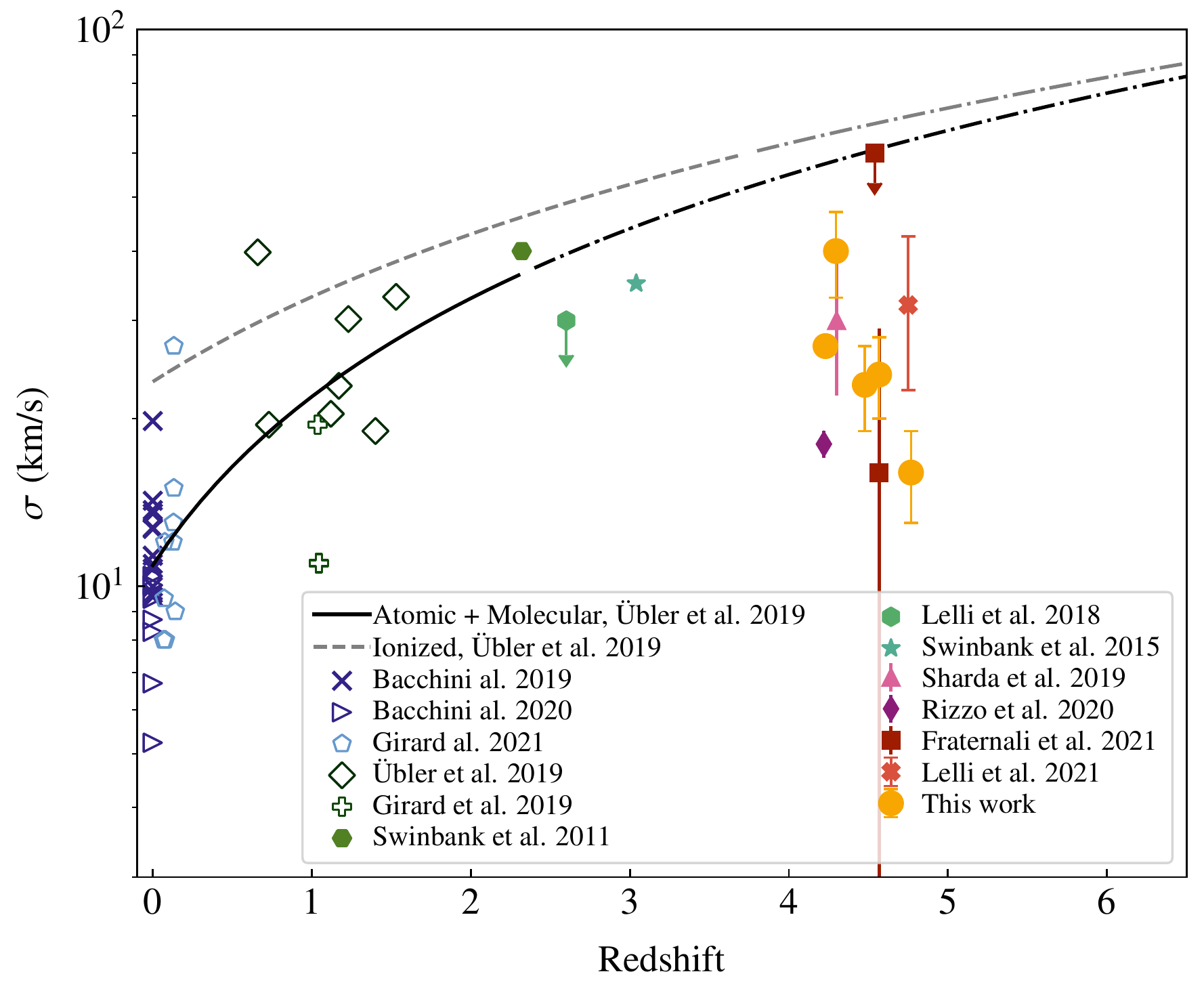}
    \caption{Location on the velocity dispersion versus redshift plane of the source galaxies in our extended sample. The values of $\sigma_{\mathrm{ext}}$ (column 6 in Table \ref{tab:globalkin} and column 4 in Table \ref{tab:sample_ext}) are shown here. The blue crosses are velocity dispersions from HI data \citep{bacchini19}. The empty markers are CO velocity dispersions  for main-sequence galaxies \citep{bacchini, girard21, ubler19, ubler18, girard}. The filled markers are velocity dispersions for DSFGs from CO transitions \citep{swinbank11, swinbank15, sharda}, [CI] \citep{lelli18} or [CII] emission lines \citep{fraternali20, rizzo3, lelli21}. The black solid and gray dotted lines show the best-fit relations to velocity dispersions measured from warm ionized tracers (H$\alpha$, [OII], [OIII]) and atomic (HI)/molecular (CO) tracers \citep{ubler19}. The dot-dashed lines show the corresponding extrapolations up to $z \approx 6$.}\label{fig:sigmaz}
\end{figure*}

Our high-resolution 3D kinematic analysis shows that the sample of DSFGs studied in this paper has ratios of $V/\sigma$ similar to those measured for spiral galaxies in the local Universe \citep{lelli16, bacchini19}. The comparison with the intermediate-$z$ star-forming galaxies is, instead, challenging, due to different gas tracers used in the literature to determine the evolution of the dynamical properties of galaxies across cosmic time. While it is firmly established that the kinematics of the molecular gas traces the galaxy kinematics, there is
an open debate on the validity of this assumption for the ionized gas tracers at high redshift \citep{girard, levy, lelli18}. For example, by comparing the kinematics of the ionized ([OIII]) and neutral gas ([CI]) for a starburst galaxy at $z = 2.6$, \citet{lelli18} concluded that [OIII] traces the outflow motions, while [CI] is a tracer of the galaxy virial motions. On the other hand, \citet{ubler18} found that the kinematics of the H$\alpha$ and CO(3-2) emission line are consistent in a star-forming galaxy at $z = 1.4$. The comparison between the dynamic properties of our DSFG sample with those found for intermediate-$z$ galaxies from ionized gas tracers, reveals that the $V/\sigma$ ratios of $z \sim 4.5$ galaxies are, on average, larger than $V/\sigma$ measured from the ionized gas at lower redshifts. This difference may be ascribed to a combination of the following reasons:
\begin{itemize}
    \item the samples of galaxies in the redshift ranges 1 to 2 with H$\alpha$ or [OII] measurements have typical rotation velocities $\lesssim$ 250 km s$^{-1}$ \citep[e.g.,][]{edt1, swinbank17, ubler17}. The DSFGs in our sample have $V_{\mathrm{flat}}$ from $\approx$ 200 to 360 km s$^{-1}$ (see Table \ref{tab:globalkin}) and a median value of 280 km s$^{-1}$. The velocity dispersions $\sigma_{\mathrm{ext}}$ of our sample (from 16 to 40 km s$^{-1}$, see Table \ref{tab:globalkin}) are consistent with the distribution of the H$\alpha$ velocity dispersions of $z \approx$ 1 star-forming galaxies found by \citet{edt1} and \citet{ubler19}, but they overlap only with the low velocity-dispersion distribution at $z \gtrsim 1.5$ \citep{ubler19}. 
    \item The velocity dispersions measured from the ionized tracers is, on average, higher than those measured from the molecular and neutral media tracers both at low- \citep{varidel, girard21} and high-$z$ \citep{girard}. 
    \item The H$\alpha$, [OIII], and [OII] emission lines are not good tracers of the galaxy dynamics. A recent study \citep{levy} on a sample of local galaxies showed, for example, that the kinematics of the molecular and ionized gas are different \citep{levy} and that the difference may be ascribed to stellar feedback processes. In particular, the measured ionized gas kinematics is affected by the presence of gas in outflows or in extraplanar layers. Unfortunately, the lack of good quality data of both the cold and warm gas for large samples of galaxies at $1 \lesssim z \lesssim 3$ prevents an accurate comparison between the kinematics of multiple phases of the ISM.
    \item The measured $V/\sigma$ values can be biased towards low values due to some residual beam-smearing effect in some studies \citep[see discussion in][]{edt1}. 
    \item The galaxies in our sample are starbursts (with the exception of SPT0113-46) while most of the galaxies with H$\alpha$ measurements at $1 \lesssim z \lesssim 2$ are main-sequence galaxies. The difference between these two galaxy populations should be taken into account before any detailed comparison of their dynamical properties. Furthermore, we do not expect an evolutionary connection between these two galaxy populations, that is, the starburst galaxies at $z \gtrsim 4$ are not the progenitors of the main-sequence galaxies at $z \lesssim 2$ (see Section \ref{sec:dsf_to_etg} for further details).
\end{itemize}
We note, also, that the $V/\sigma$ ratios measured for our DSFG sample are similar to those measured for the five galaxies of the comparison sample shown in Table \ref{tab:sample_ext} (see discussion in Section \ref{sec:bias} for the complete sample). We exclude that our results are affected by any specific assumption made to derive the kinematics of our sample \citep[see Appendix \ref{appendix:test} and ][for an extensive testing of the modelling technique used in this paper]{rizzo2}. We share, indeed, some methodologies and assumptions with previous works focusing on the kinematic study of high-$z$ galaxies \citep[e.g.,][]{swinbank17, turner, lelli18, ubler19, wisnioski20, fraternali20}. We employ, indeed, a forward modelling approach similar to \citet{edt}, \citet{lelli18}, \citet{edt18} and \citet{fraternali20}. However, while they inferred model-independent velocities and velocity dispersions, we assume specific functional forms to describe the rotation velocity and velocity dispersion profiles, similarly to \citet{swinbank17}, \citet{turner} and \citet{wisnioski20}.\\
The comparison of the velocity dispersions of galaxies at different redshifts from similar tracers (i.e., HI, CO, [CII]) indicate that there is a weak decrease of the turbulence with time. The median value\footnote{The median $\sigma$ of 25$^{+11}_{-9}$ km s$^{-1}$ is obtained by including the upper limit of 60 km s$^{-1}$ \citep[][red square in Fig. \ref{fig:sigmaz}]{fraternali20} as fiducial value. If this measurement is excluded, the resulting median $\sigma$ is 25$\pm 7$ km s$^{-1}$.} of $\approx$ 25 km s$^{-1}$ for the sample of $z \gtrsim 4$ DSFGs (markers with errorbars in Fig. \ref{fig:sigmaz}) is only a factor of $\approx$ 2 and 3 larger than the median HI $\sigma$ (12$\pm 2$ km s$^{-1}$) and CO $\sigma$ (9$^{+1}_{-2}$ km s$^{-1}$), measured in nearby galaxies \citep{bacchini19, bacchini}. The comparison with the measurements from cold gas tracers (CO, [CI]) at intermediate-$z$ indicates, instead, that the median $\sigma$ at $z \gtrsim 4$ is similar to the median $\sigma$ of 26$^{+10}_{-7}$ km s$^{-1}$, measured for $1 \lesssim z \lesssim 3$ galaxies \citep{swinbank11, swinbank15, ubler19, ubler18, lelli18, girard}. Larger samples of velocity dispersions from cold gas tracers at intermediate and high-$z$ will be needed to confirm this trend and make a robust comparison between similar tracers and galaxy populations.\\
In Fig. \ref{fig:sigmaz}, we also show the relations between velocity dispersions and redshifts derived by \citet{ubler19} for main-sequence galaxies. In particular, the gray dotted and solid black lines are the best-fit relations for velocity dispersions measured from ionized (H$\alpha$, [OII], [OIII]) and atomic/molecular tracers (HI, CO) up to $z = 3.8$ and 2.4, respectively. The dash-dotted lines show, instead, the respective extrapolations up to $z \approx 6$. The comparison between the measured velocity dispersions at $z \gtrsim 3$ and the dash-dotted lines indicates that the extrapolated $\sigma$ - $z$ relations are not valid at these high redshifts since they systematically overestimate $\sigma$ by a median factor of $\approx 2$. 
We note, also, that this mismatch is even larger if we take into account that the $\sigma$ - $z$ relations were derived mainly from normal star-forming galaxies, while both the galaxies in our sample and the four galaxies from the literature \citep{lelli18, sharda, fraternali20} are above the main-sequence and therefore more prone to develop high turbulent motions \citep[e.g.,][]{krum, hung}.

\subsection{Evolutionary path: from DSFGs to ETGs}
\label{sec:dsf_to_etg}

In this section, we compare the physical properties of the non-merging DSFGs with their plausible descendants, local ETGs. In particular, we focus on the position of these galaxies in the size-stellar mass plane ($R_\mathrm{e}$ - $M_\mathrm{star}$). The relation between the size and stellar mass of galaxies, and how it evolves with cosmic time, provides, indeed, important insights onto the assembly history of galaxies \citep[e.g.,][]{hodge16, vanderwel, lang}.

\begin{figure*}
    \centering
    \includegraphics[width=1\textwidth]{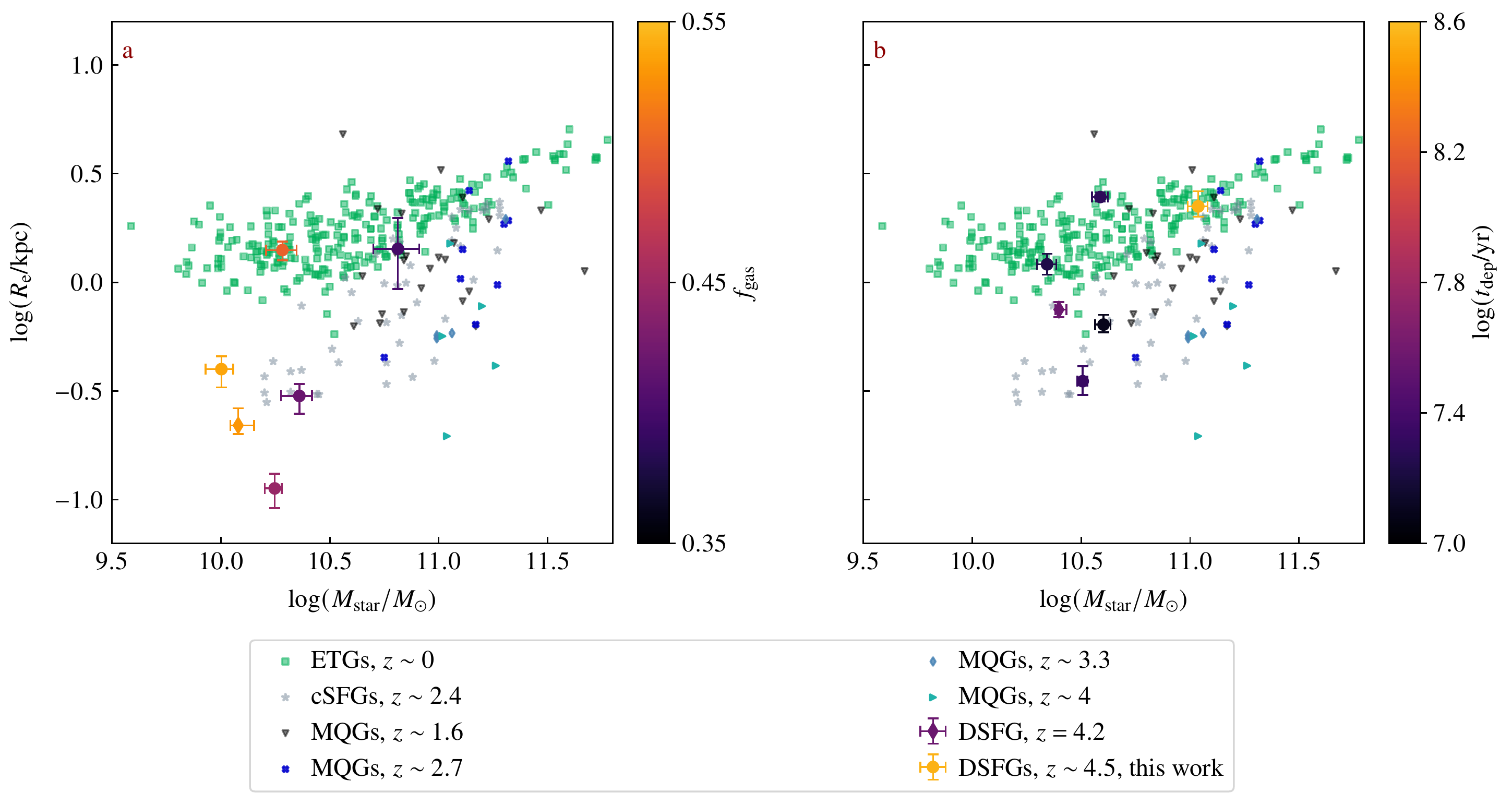}
    \caption[$R_\mathrm{e}$-$M_\mathrm{Star}$ plane]{Location on the size versus stellar mass plane for the source galaxies in our sample colour-coded according to their gas fraction (panel a, see column three of Table \ref{tab:dynamic2}) and depletion time (panel b, see column six of Table \ref{tab:dynamic2}). In panel b, the stellar masses and sizes are obtained under the assumption that all the observed gas will be converted into stars, preserving the disc configuration (see the baryonic quantities in columns four and five of Table \ref{tab:dynamic2}). Under this assumption, the sizes should be considered as upper limits. The green squares correspond to local ETGs from the ATLAS$^{\mathrm{3D}}$ survey \citep{cappellari2013a}, the gray stars are the cSFG at $z \sim$ 2.4 \citep{barro} and the black triangles, cyan crosses, blue diamonds are massive quiescent galaxies at the median redshifts indicated in the legend and they are taken from \citet{belli, lustig, esdaile, kubo}, respectively.}
    \label{fig:etg}
\end{figure*}

In Fig.~\ref{fig:etg} panels a and b, we show the position of our sample (circles) and SPT0418-47 (diamond) in the size-mass plane, colour-coded according to the gas fraction and depletion time ($M_\mathrm{gas}/$SFR), respectively (see column six of Table \ref{tab:sfr}). We note that we do not include the non-lensed galaxies listed in Table \ref{tab:sample_ext}, as the few resolution elements across their discs do not allow a rotation curve decomposition similar to that performed for the lensed sample. In Fig.~\ref{fig:etg}, we show also local ETGs \citep{cappellari2013a} and high-$z$ massive quiescent galaxies \citep{belli, lustig, esdaile, kubo}. The gray stars are $z \sim 2$ compact star-forming galaxies \citep[cSFGs][]{barro}. The cSFG have been proposed as a transition population between the star-forming and quiescent systems \citep[e.g.,][]{barro, vandokkum}. We note that for these comparison samples, only half-light radii are available in the literature, while the sizes derived from our rotation curve decomposition are the half-mass sizes. Systematic differences between these two measurements of size may arise from radial gradients in the mass-to-light ratios \citep[e.g.,][]{tortora, mosleh, suess}. For self-consistency, we therefore show in Fig.~\ref{fig:etg} the half-mass radii obtained after correcting the observed half-light radii with the relation by \citet{suess}\footnote{The half-light radii of the local ETGs are rescaled by a factor of 0.7 \citep{suess_l}. For the high-$z$ star-forming and quiescent galaxies at $z \sim$ 1 - 2 \citep{barro, belli, lustig}, the half-mass radii where obtained by multiplying the observed half-light ratio by the ratio $r$, such that $\log{r} = a[\log{M_{\mathrm{star}}-10]+b}$, with $a$ and $b$ taken at the corresponding galaxy population and redshifts from Table 2 in \citet{suess}. The resulting $r$ are of the order of $\approx$ 0.7 - 0.8. Since there are no available relations for $r$ at $z \gtrsim 3$, we applied the relation valid in the redhsift range 2.0 - 2.5 for the two samples at $z \sim 3.3$ and 4 from \citet{esdaile} and \citet{ kubo}, resulting in values of $r$ of 0.8 - 0.9.}.\\

Interestingly, SPT0113-46, the only main-sequence galaxy in our sample (see Fig.~\ref{fig:ms}), is consistent with the size-mass relation of local ETGs (panel a of Fig.~\ref{fig:etg}). Its depletion time of $\sim 357$ Myr suggests that it can rapidly consume all of its gas and become a quiescent system. There is, indeed, growing evidence of the existence of a population of quiescent galaxies at $z \gtrsim 3$ \citep[e.g.,][]{glazebrook, tanaka, valentino}. Moreover, three galaxies in our sample have Sérsic index $n \approx$ 1, indicative of a discy stellar component, three other have $n \approx$ 2, typical of discy bulges, while SPT0113-46 has the largest Sérsic index, $n \approx 6$. 
The Sérsic indices derived from our analysis suggest that the spheroidal components observed in $z \lesssim 3$ quiescent galaxies \citep{kraj, belli, lustig} were already in place when the Universe was about 10 per cent of its current age.\\

In Fig.~\ref{fig:etg} panel b, we show how the stellar mass and size of the galaxies in our sample will evolve if all of their observed gas were to be converted into stars while preserving the disc configuration. This scenario is not an impossible one: while some of the gas may be expelled by galactic outflows \citep{nelson}, more could also be accreted \citep[e.g.,][]{bouche, dekel09}. On the other hand, the sizes calculated under this assumption should be considered only as upper limits: dissipative processes lead to an accumulation of gas mainly in the central regions of galaxies, effectively shrinking the sizes of their bulges \citep{dekel, zolotov15, tacchella16}. However, it is expected that at $z \lesssim 2$, there is a strong increase of the galaxy sizes of the quiescent population, mainly driven by dry minor mergers \citep[e.g.,][]{bezanson, naab09, oser, cassata}. Panel b of Fig.~\ref{fig:etg} shows that all galaxies in our sample will end up with a stellar mass typical of the local ETGs or cSFG already at $z \approx 4$. This finding allows us to put some constraints on the physical processes \citep[e.g., mergers, accretion,][]{naab, bouche} that will be acting on these galaxies in the following $\sim$ 12 Gyr: they should either preserve the stellar mass or be responsible only for a mild growth in size.

We note that the depletion times for our sample range from $\approx$ 10 to 360 Myr (see column six of Table \ref{tab:dynamic2}), with a median value of 20 Myr. All galaxies in the sample are at $z > 4$, meaning that potentially they could be the progenitors of the recently discovered quiescent systems at $3 \lesssim z \lesssim 4$ \citep[e.g.,][]{glazebrook, tanaka, valentino}. Unfortunately, due to the lack of instruments able to spatially resolve the restframe optical/near-infrared emission at $z \gtrsim 3$, a comparison with our DSFGs on the size-stellar mass plane is currently not feasible. However, we note that in Fig.~\ref{fig:etg}, we show three samples of ten, four and five quiescent galaxies at median $z \sim 2.7$ \citep{lustig}, $z \sim 3.3$ \citep{esdaile} and $z \sim 4$ \citep{kubo}, respectively. For these two samples, the $R_\mathrm{e}$ values were derived from the restframe continuum emissions at $\approx$ 4000, 3500 and 4200 \AA, that is, the size measurements are biased towards the young stellar populations. Considering these caveats, a quantitative comparison between the high-$z$ quiescent population and our sample of DSFGs on the size-stellar mass plane is challenging. The next generation of instruments, such as the James Webb Space Telescope \citep[JWST,][]{gardner}, will allow one to measure the structural (e.g., sizes) and the stellar-population properties of these high-$z$ populations, facilitating further investigations of the evolutionary connection between the DSFGs and the $z \sim 3$ quiescent galaxies.

\section{Conclusions}
\label{sec:conclusions}

In this paper, we presented ALMA observations of the [CII] emission line for five gravitationally lensed dusty star-forming galaxies at redshift between 4 and 5. Using our lens and kinematic modelling technique, we reconstructed the background sources and inferred their kinematic and dynamical properties on $\sim$ 200-pc scales. By combining these observations and analysis with those from \citet{rizzo3} for the lensed galaxy SPT0418-47, we have thus built a sample of six DSFGs. This is the largest sample of $z \sim 4$ galaxies with such high-quality data and spatial resolutions of their [CII] line emission to date. In addition, when feasible, we also include in our sample four non-lensed DSFGs for which accurate kinematic measurements were derived \citep{sharda, fraternali20, lelli21}.

The measurement of the rotation velocities and velocity dispersion profiles allowed us to gain insights on how the extreme astrophysical processes and conditions characterising the early Universe affect the properties of these young galaxies.
In particular, the sample studied in this paper allowed us to confirm a previous finding \citep{rizzo3}: DSFGs have $V/\sigma$ in the range 7 to 15 and median velocity dispersion in the range between $\approx 30$ and $\approx 60$ km s$^{-1}$. To date, such dynamically cold galaxies with the measured values of SFR and gas fractions are not predicted by any model \citep[e.g.,][]{pillepich, hung, dekel20}. By investigating the velocity dispersions and SFR of the studied galaxies, we found that stellar feedback mechanisms are able to sustain the level of observed turbulence with low efficiency. There is, therefore, no need of additional drivers of turbulence, such as gravitational instabilities. We then compare the velocity dispersions in our sample with the analogue measurements from the literature for galaxies at lower redshifts and with the empirical $\sigma$ - $z$ relation found by \citet{ubler19} for normal main-sequence galaxies up to $z \lesssim 3$. The median values of the velocity dispersions in our sample is only a factor of $\approx 2$ larger than the typical HI velocity dispersions of local spiral galaxies and consistent with the CO velocity dispersions found in star-forming galaxies at 1 $\lesssim z \lesssim$ 2. We find that the extrapolation of the \citet{ubler19} $\sigma$-$z$ relation to the redshifts studied in this paper results in a systematic overestimation of the observed velocity dispersions by a median factor of $\approx 2$.

The rotation curves of the galaxies analysed here have shapes similar to those of local spirals: they flatten at large radii and show a variety of behaviours in the inner regions, from slow to steeply rising. This similarity implies that the dynamical structures of local galaxies were already in place at higher redshifts. On the other hand, the individuality of the observed rotation curves can be explained by the difference in the dynamical parameters defining the stellar, gas, and dark matter halo distributions. \\
The lack of spatially-resolved optical/near-infrared data prevented us from making sophisticated assumptions on the stellar distribution. We thus decomposed the rotation curve of our galaxies using a single Sérsic profile for the stellar component, an NFW for the dark matter halo, and an exponential disc for the gas, as traced by the [CII] line emission. From the dynamical fitting, we found that the galaxies in our sample have a stellar-mass between $\sim 1 \times 10^{10} M_{\odot}$ and $\sim 7 \times 10^{10} M_{\odot}$. Their gas fraction ranges between 0.4 and 0.6. Furthermore, four out of six galaxies have a  stellar component which is well described by a Sérsic index of $n \gtrsim 2$.\\
This analysis allowed us to put constraints on the mechanisms responsible for the transformation of these galaxies into their most plausible descendants, the quiescent systems observed at $z \lesssim 3$. In particular, by comparing our sample with local ETGs and $z\sim2$ cSFGs in the size-stellar mass plane, we found that two galaxies in our sample, SPT0113-46 and SPT2132-58, have sizes and stellar masses consistent with the low-$z$ samples, two galaxies are consistent with the low-mass end of the cSFGs, while the others cover the low-mass, small-size region. Interestingly, SPT0113-46 has the lowest gas fraction within our sample and, is the only galaxy on the main sequence. All these properties seem to be indicating that SPT0113-46 is in the process of consuming its residual gas, quenching its star formation and transforming into a typical ETG.
We also found that the baryonic masses in our sample are all consistent with those of local ETGs. This result allowed us to set constraints on the small amount of baryonic matter that can be accreted in the following $\sim$ 12 Gyr of the lifetime of these galaxies.

Our results are based on just ten galaxies. While statistically significant conclusions can not be drawn, these first results are promising. In the near future, the thousands of strong gravitationally lensed galaxies \citep{oguri, collett, mckean} discovered by
the Euclid space telescope\footnote{\href{https://www.euclid-ec.org}{https://www.euclid-ec.org}}, the Rubin Observatory\footnote{\href{https://www.lsst.org}{https://www.lsst.org}} and the Square Kilometer Array\footnote{\href{https://www.skatelescope.org}{https://www.skatelescope.org}}, combined with the capability of ALMA and JWST will provide us with the opportunity to fully characterise the structural and dynamical properties on large sample of galaxies up to the epoch of reionization.

\section*{Acknowledgements}
We thank the anonymous referee for a careful reading and helpful feedback. FR is grateful to Cecilia Bacchini and Francesco Valentino for useful comments and discussions and to Federico Lelli for providing the kinematic data on ALESS073.1. We also acknowledge the support of Mahsa Kohandel, Andrea Ferrara, Andrea Pallottini and Livia Vallini.\\
This project has received funding from the European Union’s Horizon 2020 research and innovation program under the Marie Sklodowska-Curie grant agreement No. 847523 ‘INTERACTIONS’.
SV thanks the Max Planck Society for support through a Max Planck Lise Meitner Group, and acknowledges funding from the European Research Council (ERC) under the European Union’s Horizon 2020 research and innovation programme (LEDA: grant agreement No 758853). FF acknowledges support from the Friedrich Wilhelm Bessel Research Award Programme of the Alexander von Humboldt. \\
This research made use of Astropy and Matplotlib packages for Python \citep{astropy}.




\section*{Data Availability}
The data underlying this article are available in the ALMA Science Archive, at https://almascience.eso.org/asax/. This work makes use of the following ALMA data: ADS/JAO.ALMA\#2016.1.01499.S. ALMA is a partnership of ESO (representing its member states), NSF (USA) and NINS (Japan), together with NRC (Canada), NSC and ASIAA (Taiwan), and KASI (Republic of Korea), in cooperation with the Republic of Chile. The Joint ALMA Observatory is operated by ESO, AUI/NRAO and NAOJ.

\bibliographystyle{mnras}
\bibliography{ms.bib} 



\appendix
\section{Tests on the kinematic hyperprior}\label{appendix:test}
\begin{figure}
    \centering
    \includegraphics[width=\columnwidth]{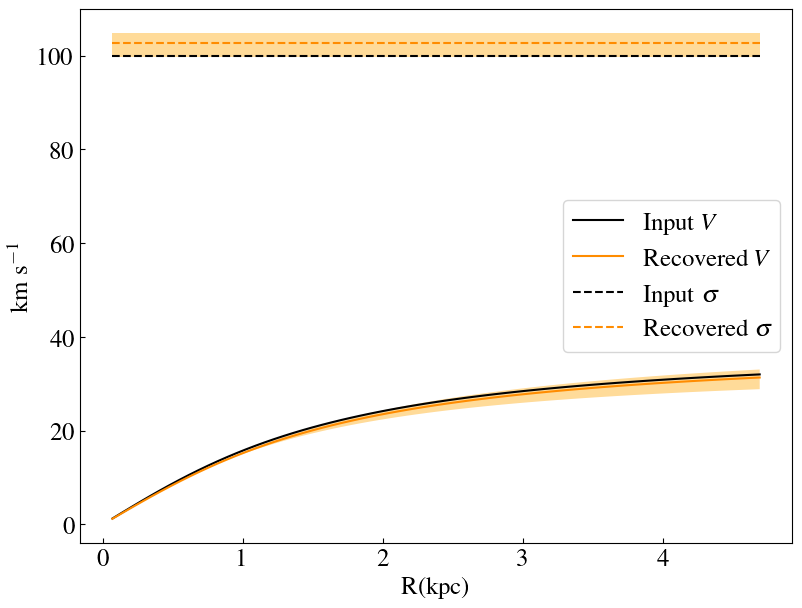}
    \caption{The black solid and dashed lines show the input profiles of the rotation velocity and velocity dispersion profiles used to create mock data of a lensed dispersion-dominated galaxy. The orange solid and dashed lines show the recovered profiles and the corresponding uncertainties. }
    \label{fig:dd}
\end{figure}

\begin{figure*}
    \centering
    \includegraphics[width=0.95\textwidth]{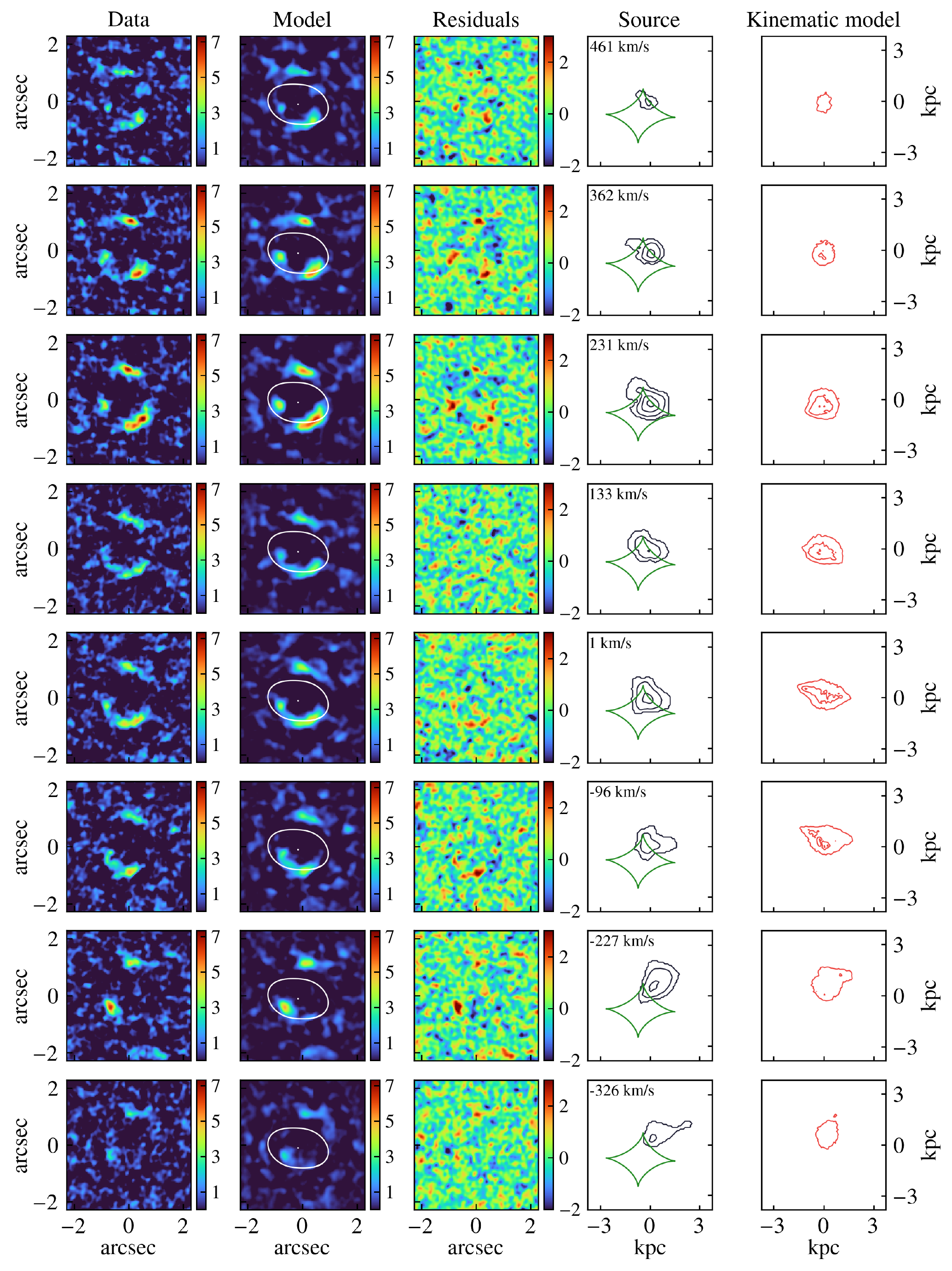}
    \caption[Channel maps for SPT0346-52]{Channel maps for SPT0346-52. Same as in Fig.~\ref{fig:03}.}
    \label{fig:spt0346_ch}
\end{figure*}

\begin{figure*}
    \centering
    \includegraphics[width=0.95\textwidth]{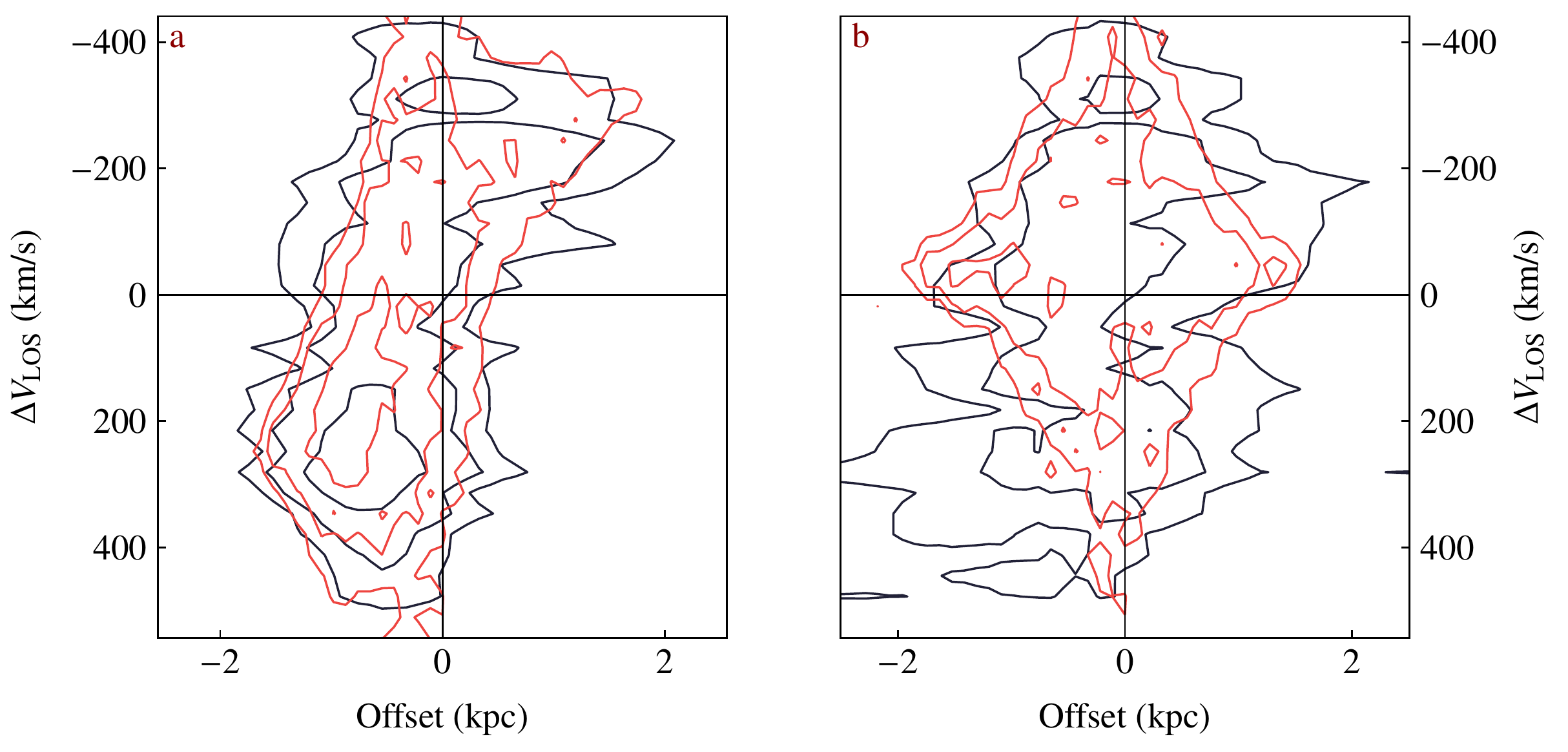}
    \caption[Channel maps for SPT0346-52]{Position-velocity diagrams for SPT0346-52. Same as in Fig.~\ref{fig:02}.}
    \label{fig:spt0346_pv}
\end{figure*}

The methodology used to derive the kinematics of the lensed galaxies was described \citet{rizzo2}, where also extensive tests on mock data containing lensed galaxies with different kinematic and geometric properties were discussed. To summarize, with this methodology, one is able to derive the rotation velocity and velocity dispersion as well as the geometry of lensed galaxies. For the latter, \citet{rizzo2} performed tests on galaxies with a variety of inclinations, from $i \sim 40$ deg to $i \sim 80$ deg (nearly edge-on), as well as on galaxies with warps in their position angles. The mock data analyzed in \citet{rizzo2} have also a variety of $V/\sigma$, from $\sim 2$ up to 10.\\
We have now extended the tests in \citet{rizzo2} to mock data containing a lensed dispersion-dominated galaxy with $V/\sigma \sim 0.3$ and to real observations of a lensed merging system. In the first case, we found the residuals are at the noise level and the input parameters are recovered with accuracies of the order of 2\%, consistent with the values reported in \citet{rizzo2}. We note that in the literature, there also are a number of non-lensed galaxies that are fitted using a tilted ring model and with resulting $V/\sigma \lesssim 2$ \citep{iorio, mancera}. In other words, a galaxy well described by a rotating disc is not necessarily a galaxy with high $V/\sigma$ ratio.  (Fig.~\ref{fig:dd})
In the second case, we model the ALMA [CII] observations of SPT0346-52, a DSFG that was previously identified as a close-in merger \citep{litke}. Consistently with this previous finding \citep{litke}, we found that a rotating disc is not a good description of this DSFG for the following reasons:
\begin{itemize}
    \item the residuals are at the noise level when we model the data without the inclusion of a kinematic hyperprior on the source. Systematic residuals up to the 5-$\sigma$ levels appear, instead, when the source is reconstructed using a rotating disc as regularising prior for the source reconstruction (see red spots in column three in Fig.~\ref{fig:spt0346_ch});
    \item the p-v diagram along the minor axis show features that are not reproduced by a rotating disc (Fig.~\ref{fig:spt0346_pv});
    \item we infer a rotation velocity of $\sim 700$ km s$^{-1}$. This value is unphysical and it is a clear hint that SPT0346-52 is a merging system.
\end{itemize}
Overall, this test shows that, even if we use a rotating disc as a hyperprior for the source reconstruction, we are able to distinguish whether a rotating disc is not a good description of the lensed galaxy by comparing the source reconstruction and the residuals obtained with and without the source hyperprior. We have also performed tests on mock observations from the SERRA simulation \citep{pallottini}, which confirm the above results and will be presented in a follow-up paper.

\section{Lens and kinematic models}
\label{appendixb}

\begin{figure*}
    \centering
    \includegraphics[width=\textwidth]{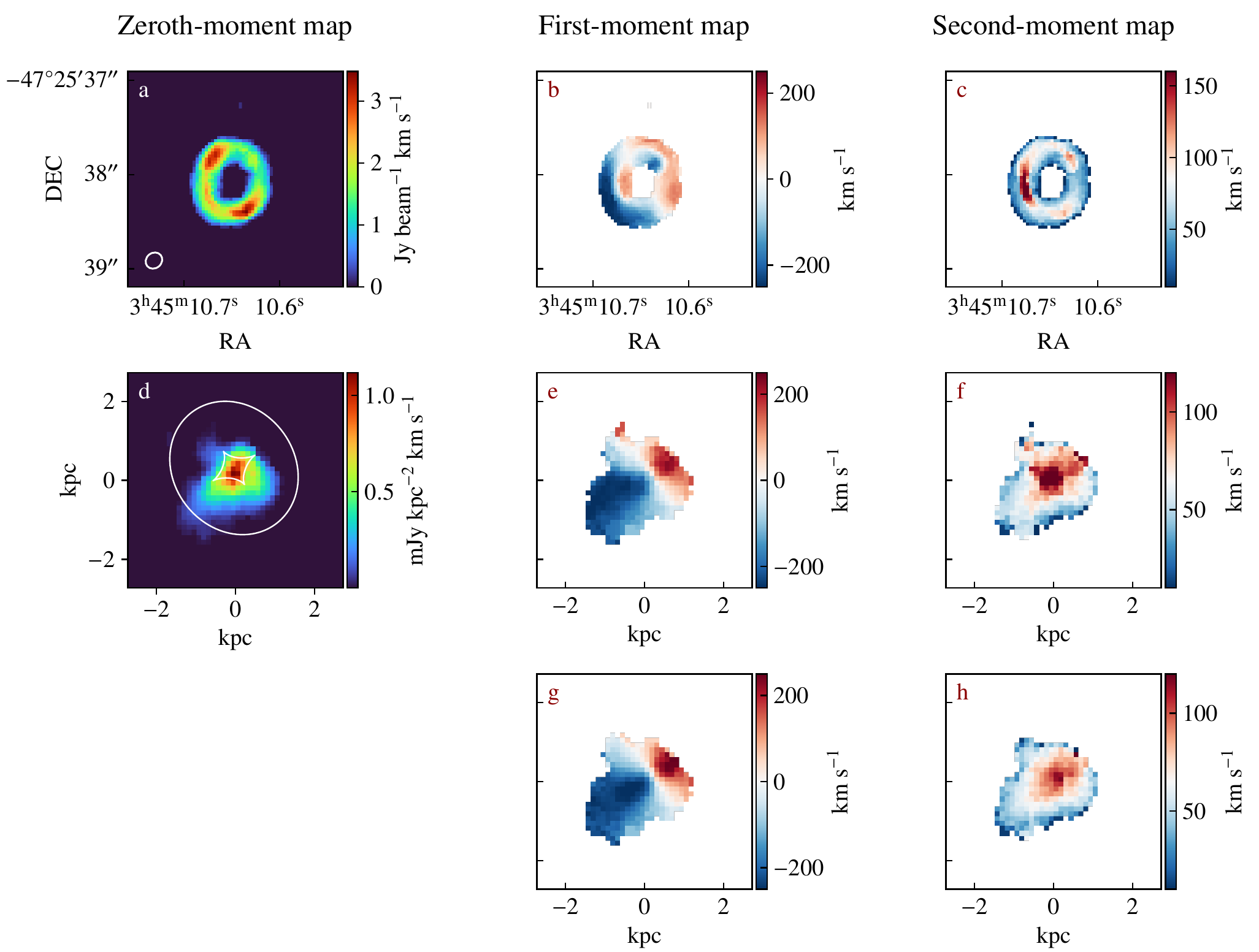}
    \caption[Moment maps for SPT0345-47]{Moment maps for SPT0345-47. Same as in Fig.~\ref{fig:01}. The beam size, shown as a white ellipse on the lower left corner of panel a, is $0.18 \times 0.16$ arcsec$^{2}$ at a position angle of -50.6$^{\circ}$.}
    \label{fig:ap4}
\end{figure*}

\begin{figure*}
    \centering
    \includegraphics[width=0.95\textwidth]{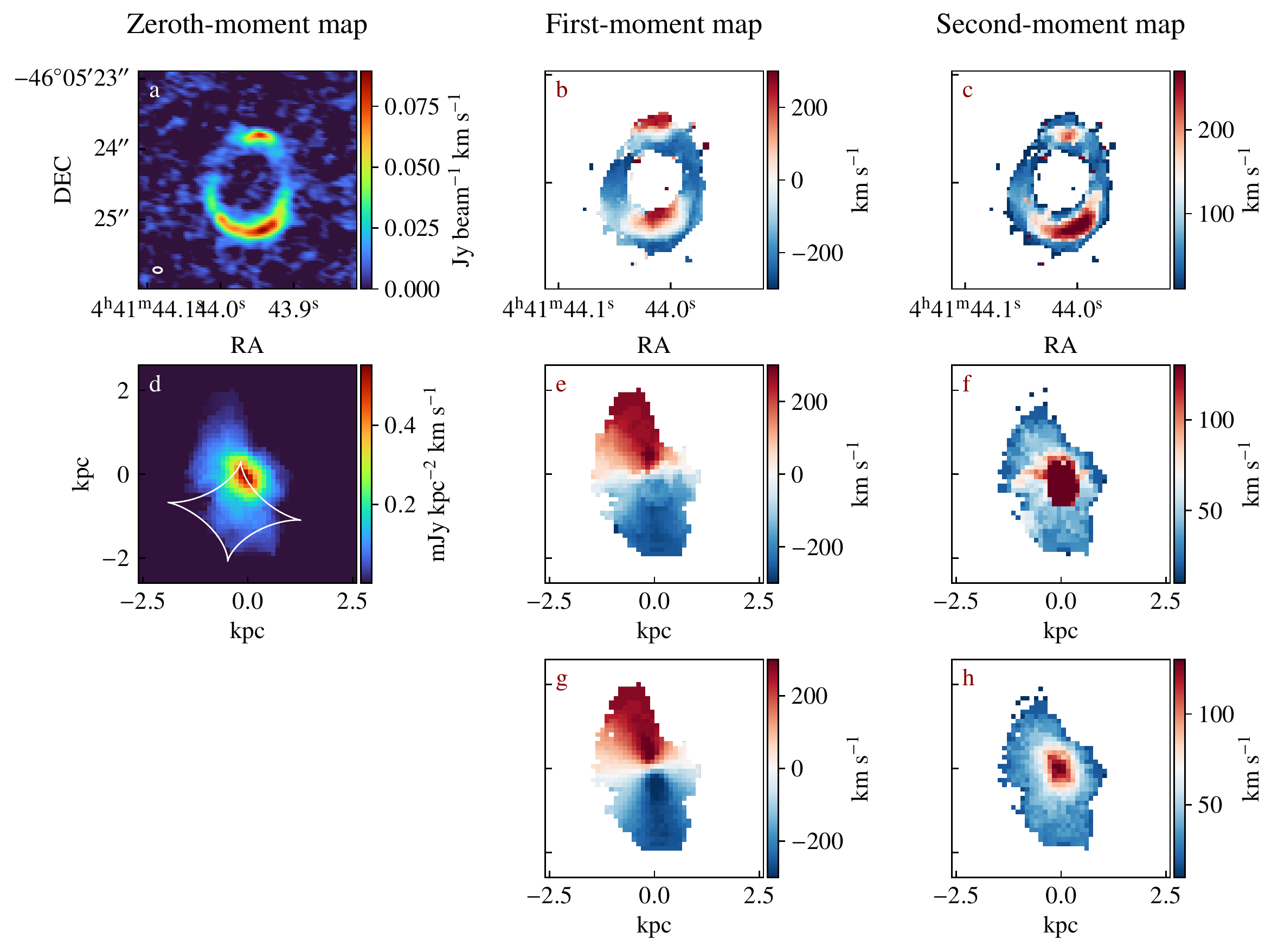}
    \caption[Moment maps for SPT0441-46]{Moment maps for SPT0441-46. Same as in Fig.~\ref{fig:01}. The beam size, shown as a white ellipse on the lower left corner of panel a, is $0.23 \times 0.19$ arcsec$^{2}$ at a position angle of -46.6$^{\circ}$.}
    \label{fig:ap7}
\end{figure*}

\begin{figure*}
    \centering
    \includegraphics[width=0.95\textwidth]{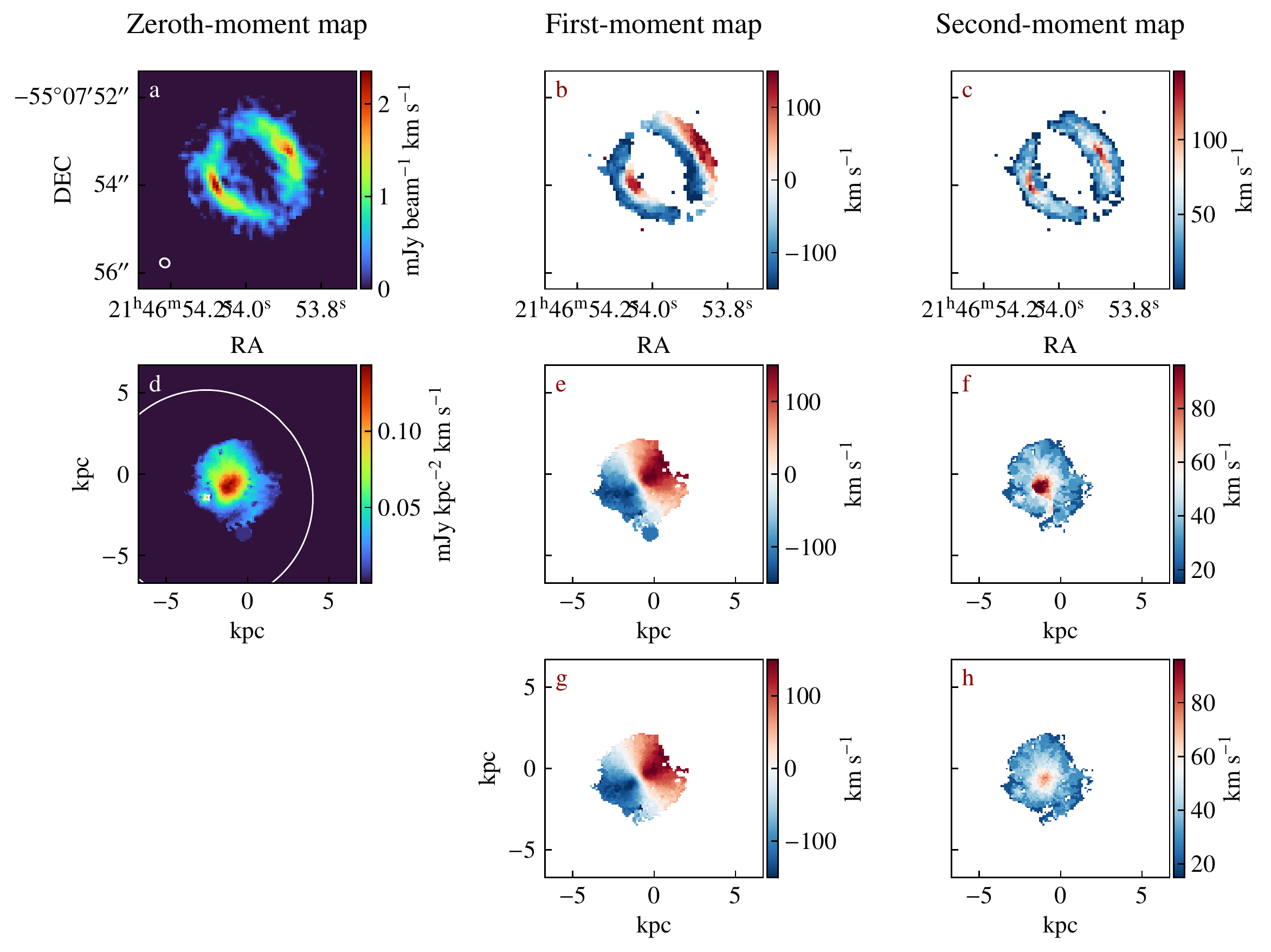}
    \caption[Moment maps for SPT2146-56]{Moment maps for SPT2146-55. Same as in Fig.~\ref{fig:01}. The beam size, shown as a white ellipse on the lower left corner of panel a, is $0.23 \times 0.20$ arcsec$^{2}$ at a position angle of -64.1$^{\circ}$.}
    \label{fig:ap10}
\end{figure*}

\begin{figure*}
    \centering
    \includegraphics[width=0.95\textwidth]{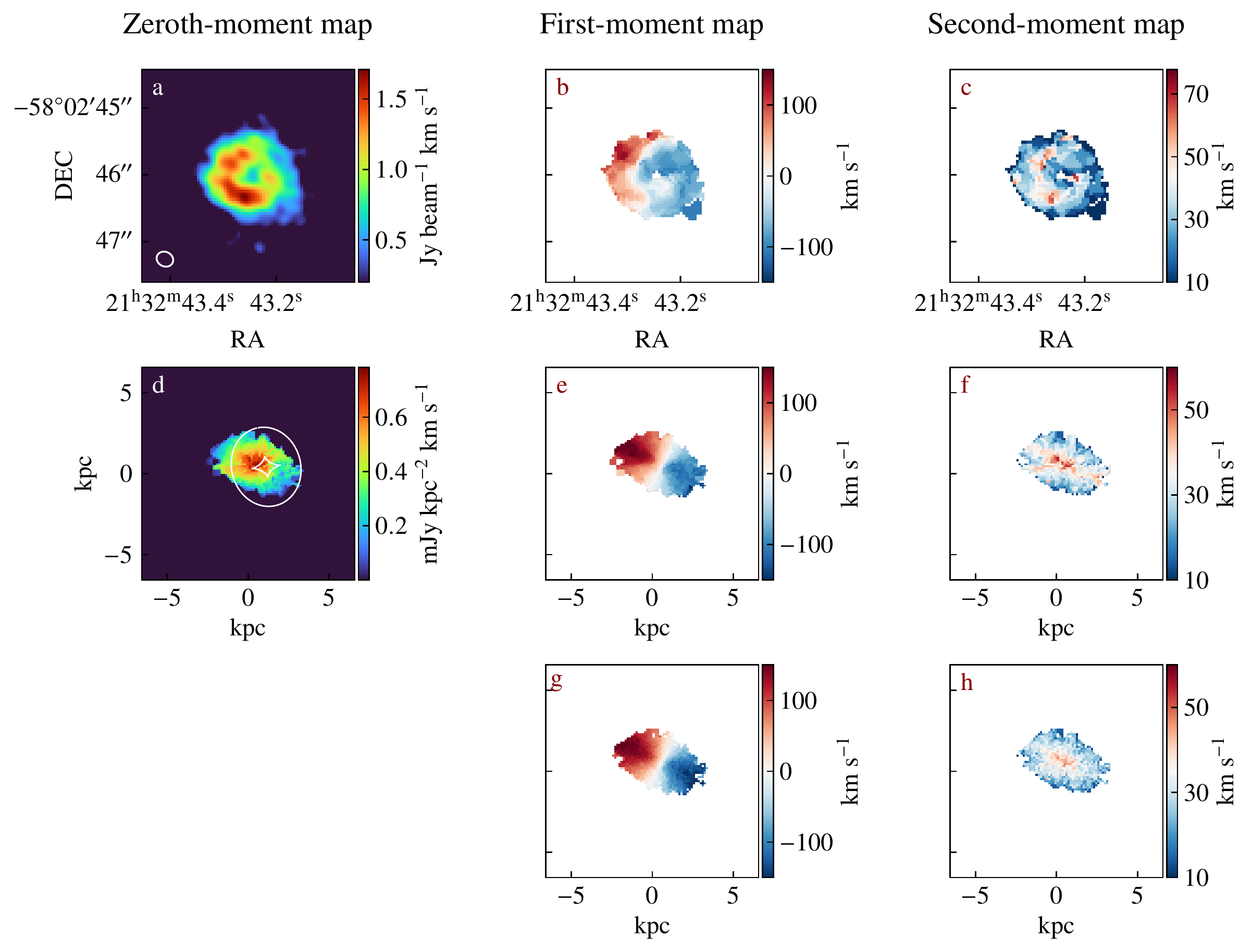}
    \caption[Moment maps for SPT2132-58]{Moment maps for SPT2132-58. Same as in Fig.~\ref{fig:01}. The beam size, shown as a white ellipse on the lower left corner of panel a, is $0.25 \times 0.22$ arcsec$^{2}$ at a position angle of 63.3$^{\circ}$.}
    \label{fig:ap13}
\end{figure*}

\begin{figure*}
    \centering
    \includegraphics[width=0.95\textwidth]{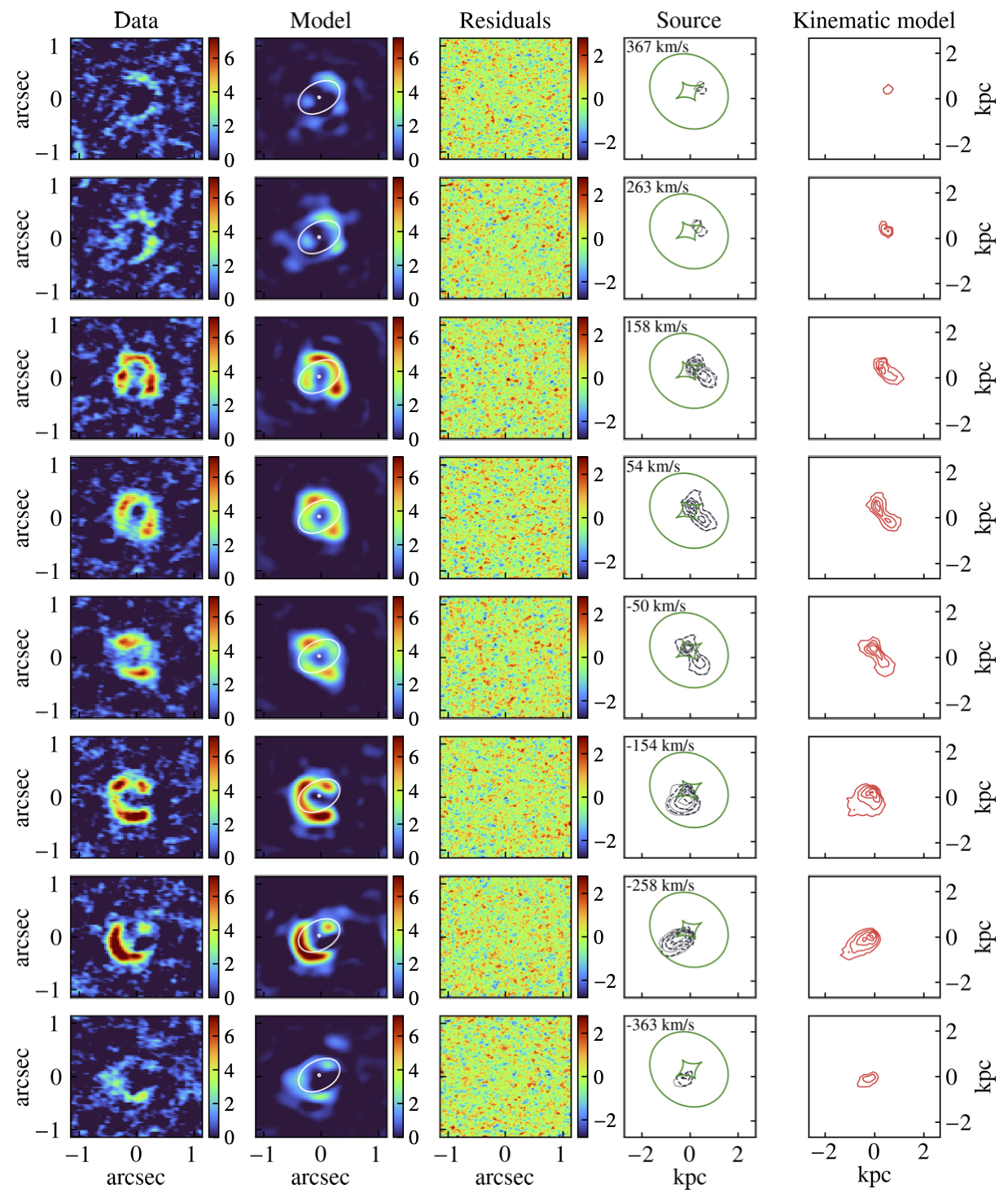}
    \caption[Channel maps for SPT0345-47]{Channel maps for SPT0345-47. Same as in Fig.~\ref{fig:03}.}
    \label{fig:ap6}
\end{figure*}

\begin{figure*}
    \centering
    \includegraphics[width=0.95\textwidth]{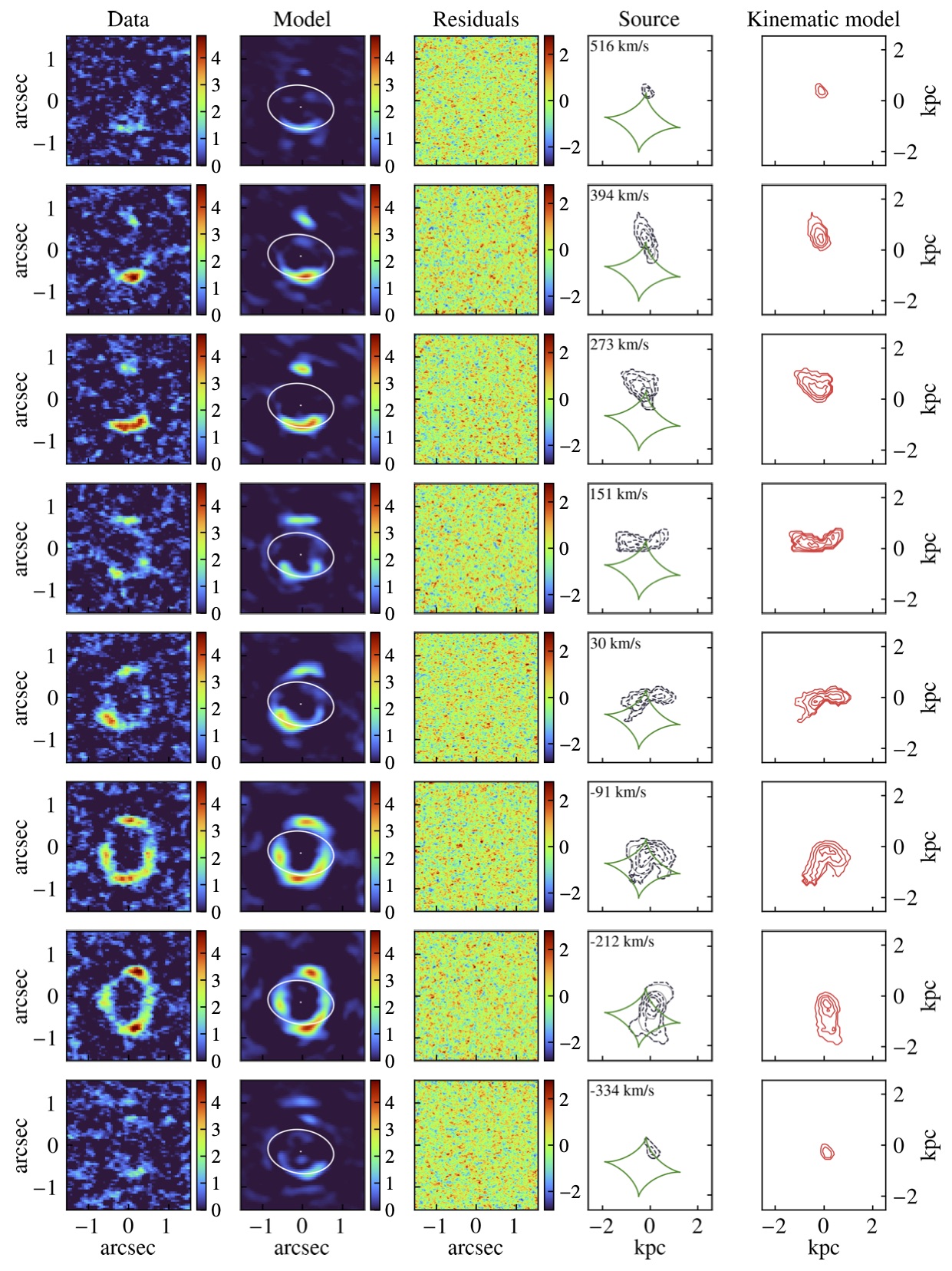}
    \caption[Channel maps for SPT0441-46]{Channel maps for SPT0441-46. Same as in Fig.~\ref{fig:03}.}
    \label{fig:ap9}
\end{figure*}

\begin{figure*}
    \centering
    \includegraphics[width=0.95\textwidth]{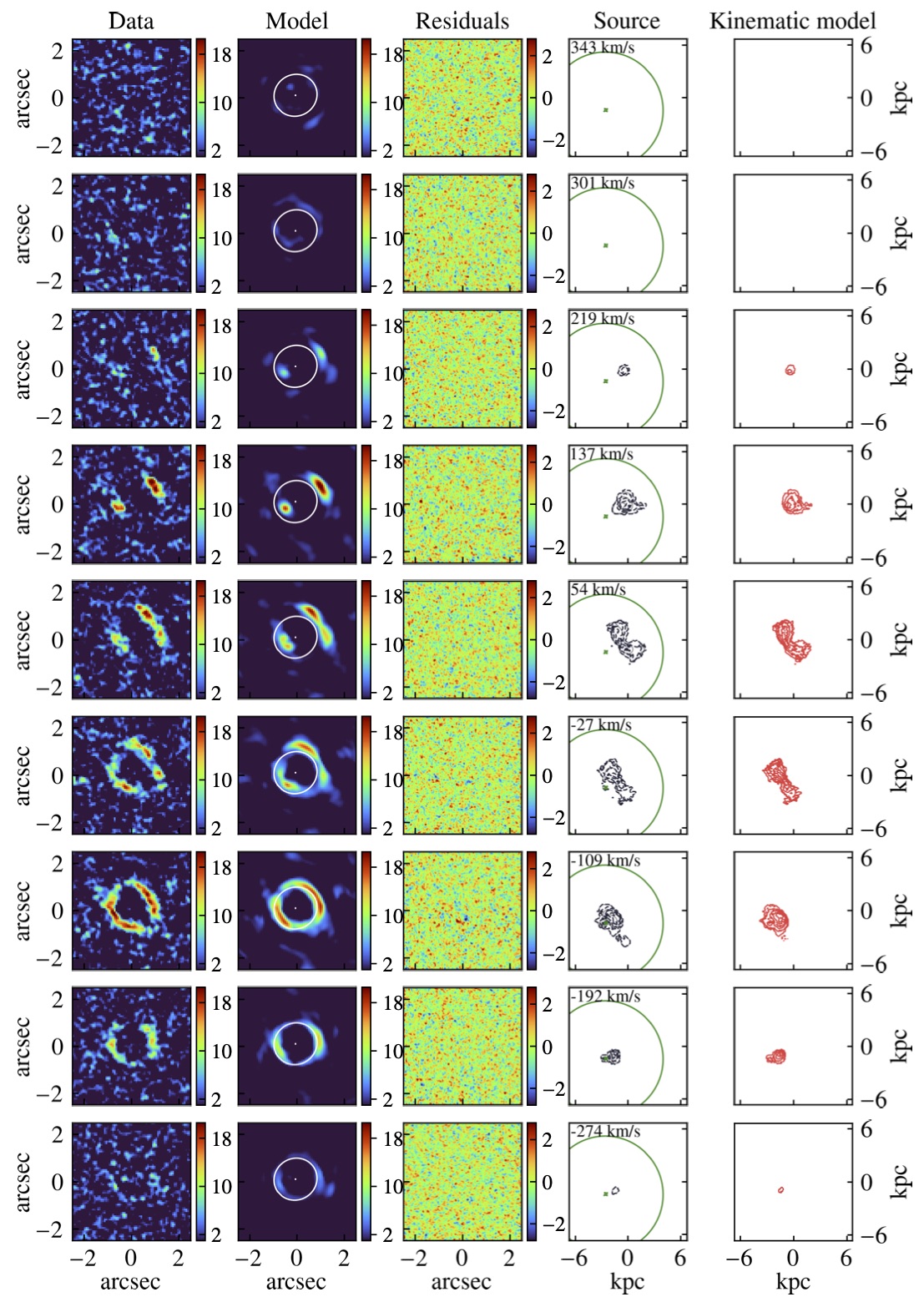}
    \caption[Channel maps for SPT2146-56]{Channel maps for SPT2146-55. Same as in Fig.~\ref{fig:03}.}
    \label{fig:ap12}
\end{figure*}

\begin{figure*}
    \centering
    \includegraphics[width=0.95\textwidth]{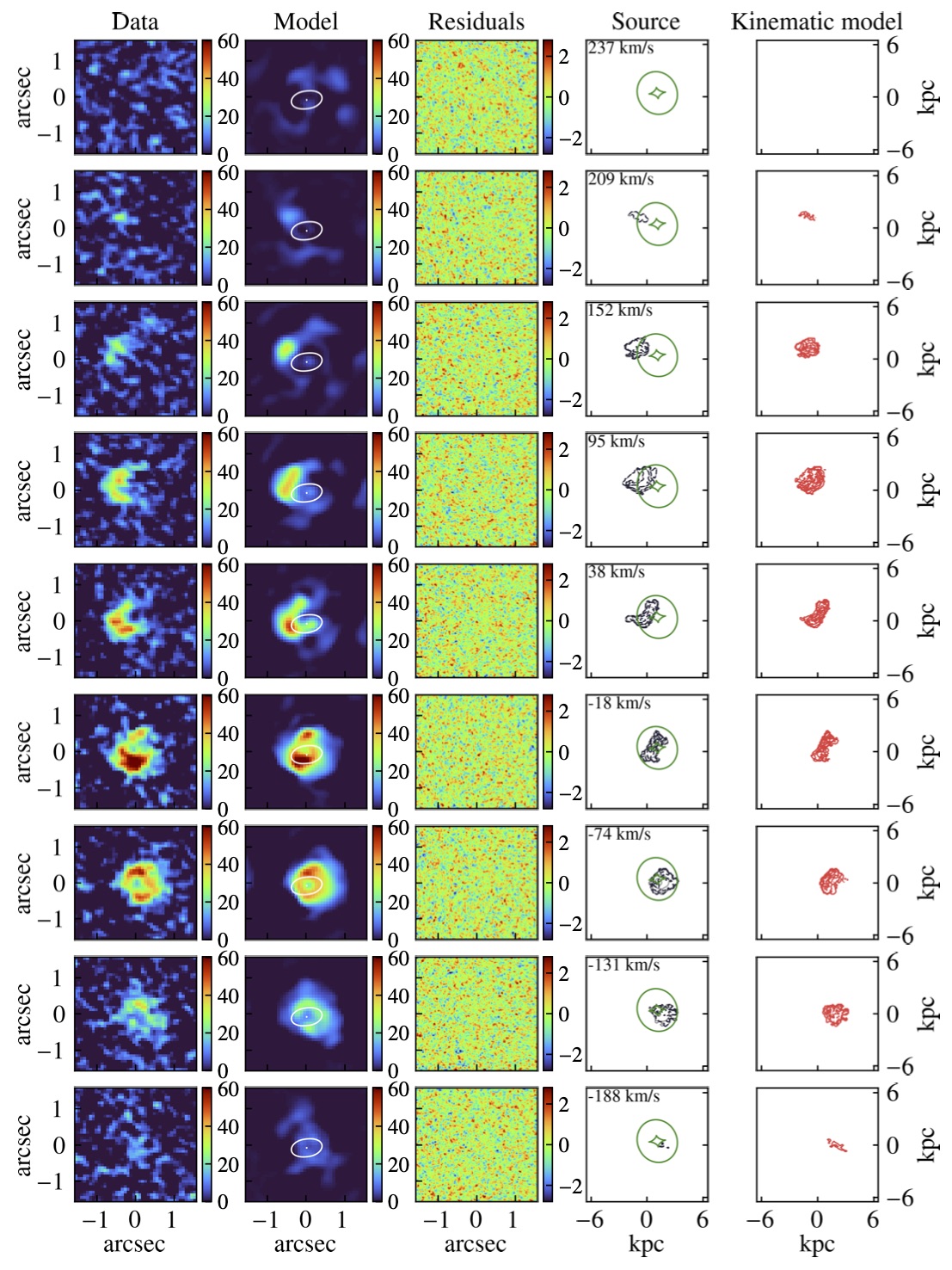}
    \caption[Channel maps for SPT2132-58]{Channel maps for SPT2132-58. Same as in Fig.~\ref{fig:03}.}
    \label{fig:ap15}
\end{figure*}

\begin{figure*}
    \centering
    \includegraphics[width=0.95\textwidth]{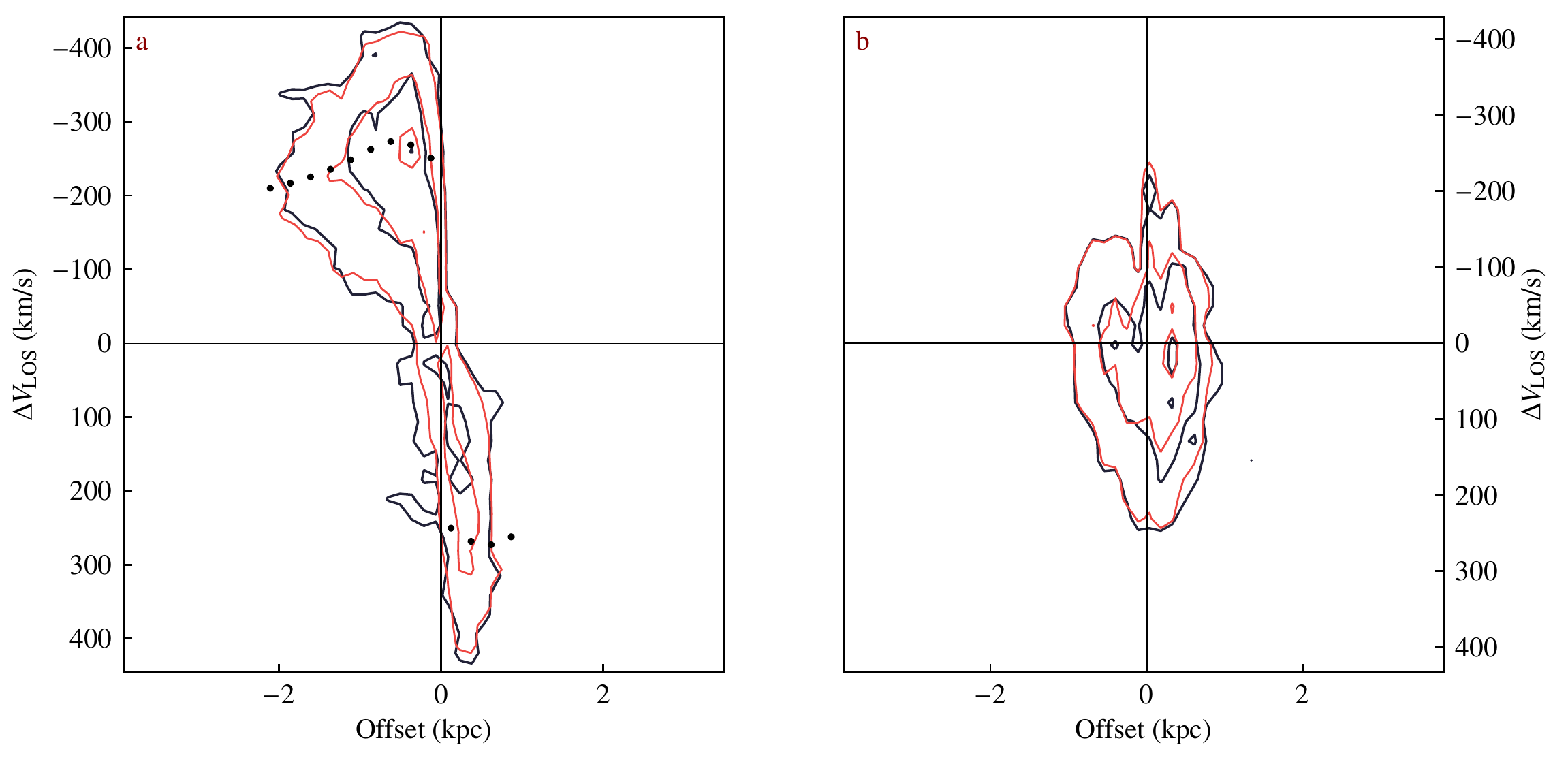}
    \caption[Position-velocity diagrams for SPT0345-47]{Position-velocity diagrams for SPT0345-47. Same as in Fig.~\ref{fig:02}.}
    \label{fig:ap5}
\end{figure*}

\begin{figure*}
    \centering
    \includegraphics[width=0.95\textwidth]{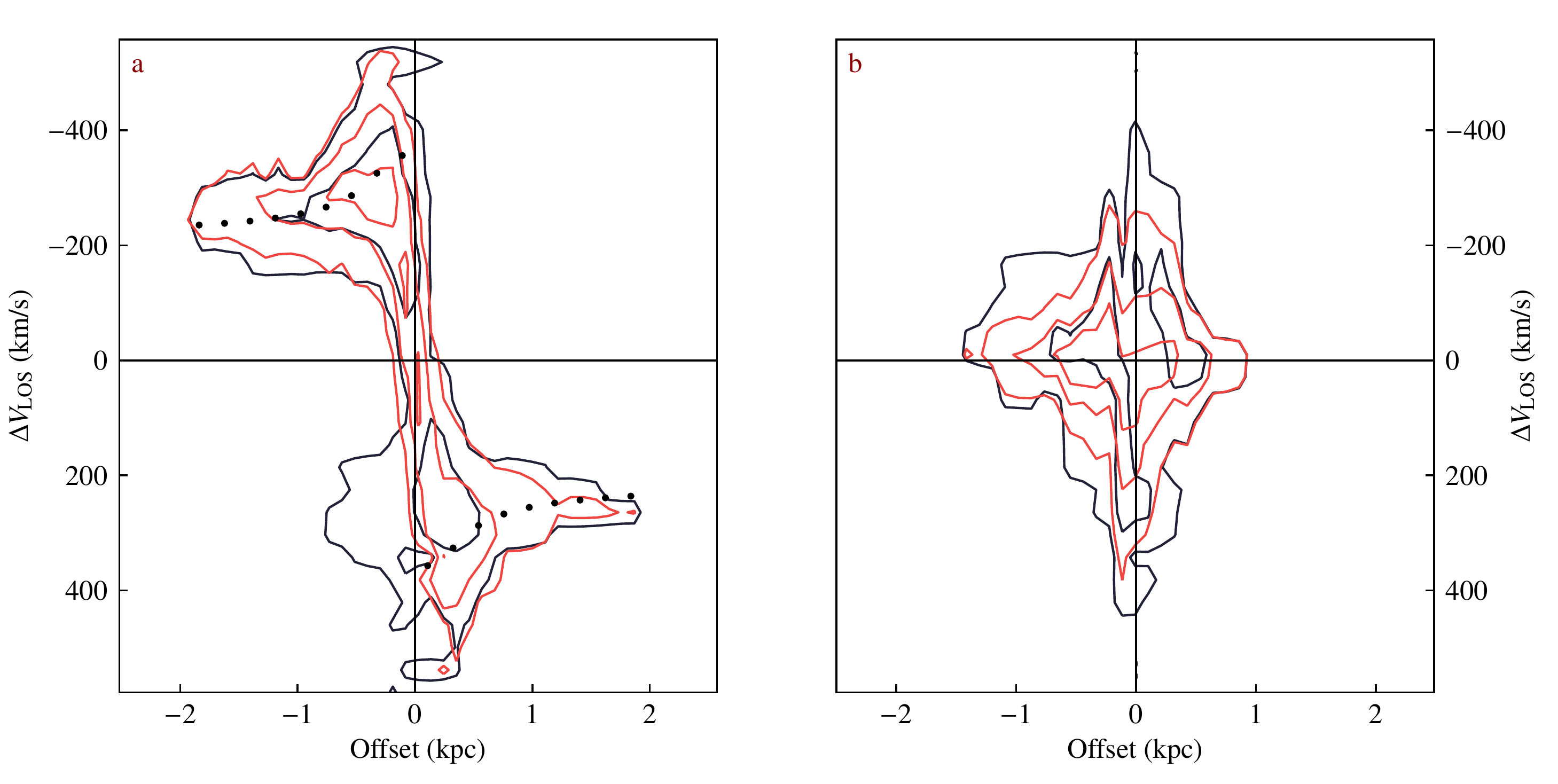}
    \caption[Position-velocity diagrams for SPT0441-46]{Position-velocity diagrams for SPT0441-46. Same as in Fig.~\ref{fig:02}.}
    \label{fig:ap8}
\end{figure*}

\begin{figure*}
    \centering
    \includegraphics[width=0.95\textwidth]{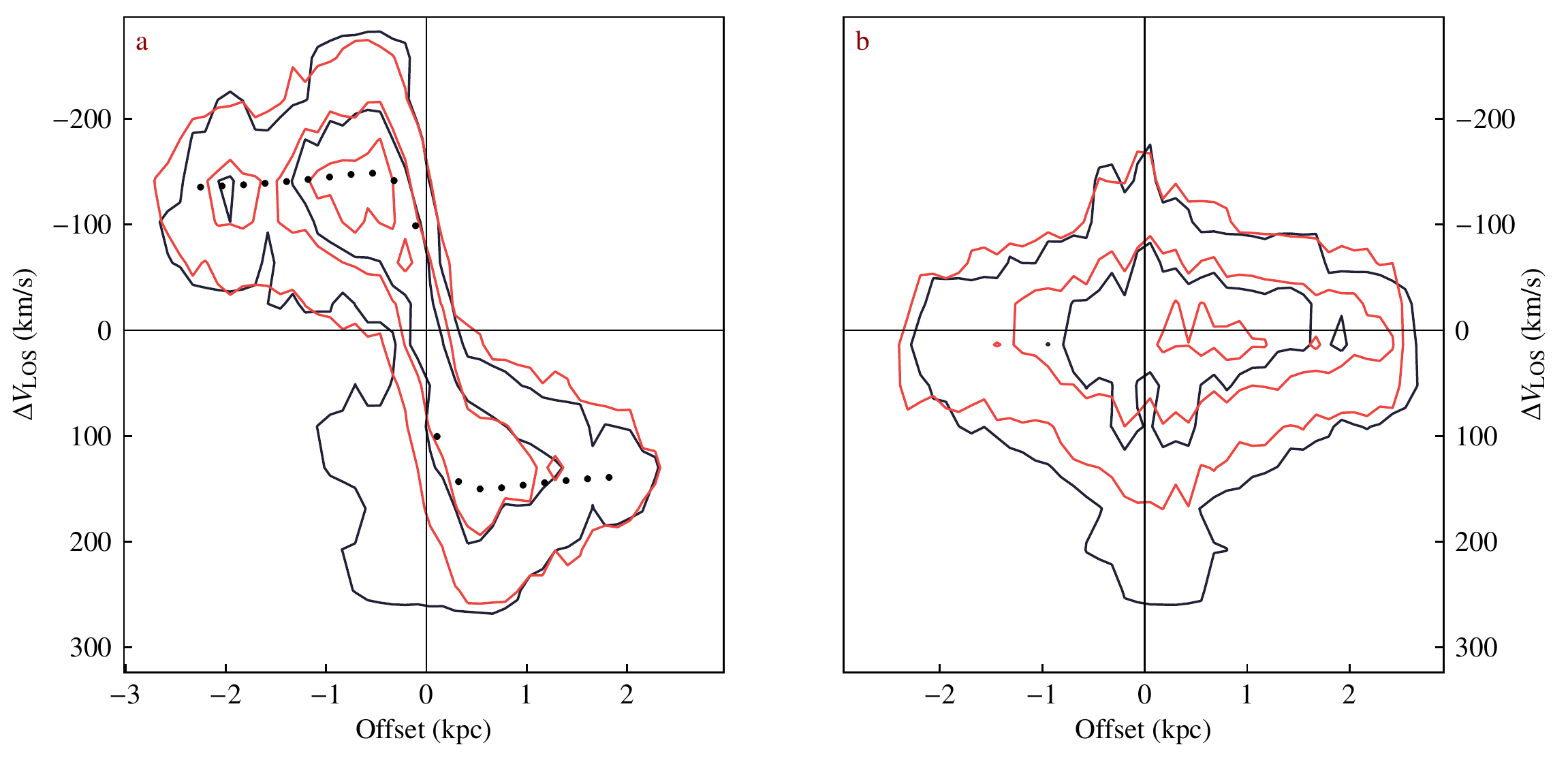}
    \caption[Position-velocity diagrams for SPT2146-56]{Position-velocity diagrams for SPT2146-55. Same as in Fig.~\ref{fig:02}.}
    \label{fig:ap11}
\end{figure*}

\begin{figure*}
    \centering
    \includegraphics[width=0.95\textwidth]{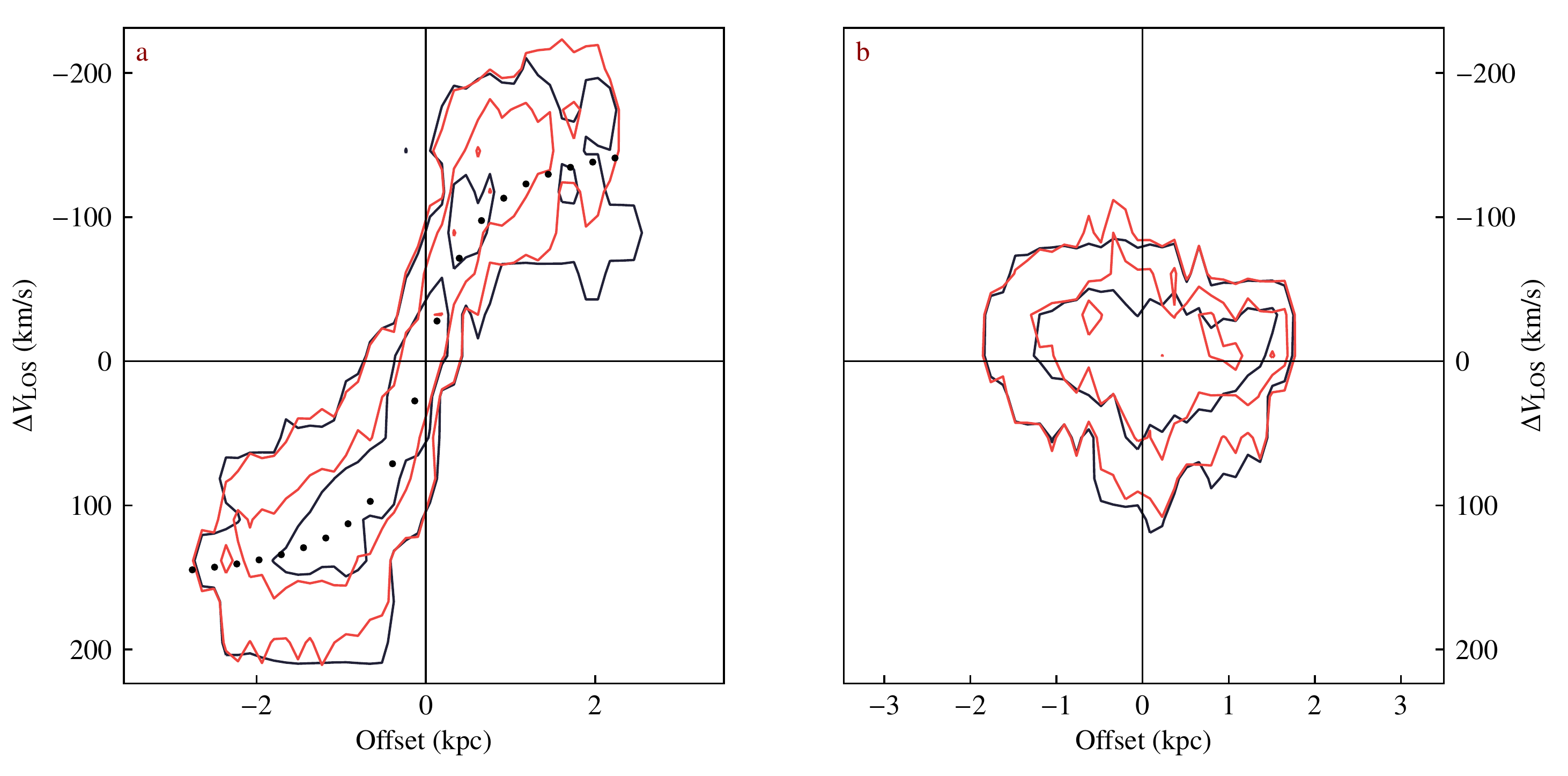}
    \caption[Position-velocity diagrams for SPT2132-58]{Position-velocity diagrams for SPT2132-58. Same as in Fig.~\ref{fig:02}.}
    \label{fig:ap14}
\end{figure*}

In this section, we show the outputs of the lens and kinematic for each galaxy of the sample. As for SPT0113-46 in Section \ref{sec:src_kin} (Fig.~\ref{fig:01} - \ref{fig:02}), we show three sets of figures: the moment map of the lensed galaxies, the corresponding reconstructed source and kinematic model (Figs.~\ref{fig:ap4}-\ref{fig:ap13}); some representative channel maps from the cubes containing the data, the model, the residuals, the source and the kinematic model (Figs.~\ref{fig:ap6}-\ref{fig:ap15}); the position-velocity diagrams along the minor and major axes (Figs.~\ref{fig:ap5}-\ref{fig:ap14}).

\subsection{Spatial resolution on the source plane}
\label{appendixc}
In order to estimate the spatial resolutions on the source plane we calculate the dimensions of the triangular grid \citep{vegetti09, rizzo2} in regions with SNR $\gtrsim 3$ for each spectral channel of the source cube (see Section \ref{sec:src_kin}). We then verified that the recovered resolutions correspond to those obtained using the methodology described in \citet{stacey20}. In this case, mock lensed data were created from Gaussian sources with different sizes and different locations on the source plane. The recovered source dimensions are then compared to the input Gaussian sources to infer the minimum reliable source sizes.\\
The minimum and median spatial resolutions for each galaxy of the sample are listed in Table \ref{tab:resolution}.  

\begin{table}
	\centering
	\caption{Statistics of the spatial resolution on the source plane.}
	\label{tab:resolution}
	\begin{tabular}{ccc} 
		\hline
		\noalign{\smallskip}
		Name & Maximum resolution & Median resolution\\
		 & pc & pc\\
	    \noalign{\smallskip}
		\hline
		\noalign{\smallskip}
		 SPT0113-46 & 26 & 198\\\noalign{\vspace{1pt}}
		 SPT0345-47 & 47 & 189\\\noalign{\vspace{1pt}}
		 SPT0441-46 & 139 & 185 \\\noalign{\vspace{1pt}}
		 SPT2146-55 & 24 & 173\\\noalign{\vspace{1pt}}
		SPT2132-58 & 64 & 300\\ \noalign{\vspace{1pt}}
    \noalign{\smallskip}
    \hline
    \noalign{\medskip}
	\end{tabular}
\end{table}

\section{Prior distributions for the dynamical modelling}
\label{appendixa}
In Table \ref{tab:c}, we show the values of the concentration parameters $c$ that define the distribution of the NFW dark-matter halo, equation (\ref{eq:vdm}). Table \ref{tab:prior} lists the intervals for the uniform and log-uniform (for the masses) priors employed in the dynamic fitting described in Section \ref{sec:src_dyn}. Similarly to the assumption in \citet{rizzo3}, for $\alpha_{\mathrm{[CII}]}$ we employ a uniform prior in the range corresponding to $\pm$ 3-$\sigma$ respect to the mean value of 30 $M_{\odot}/L_{\odot}$ found by \citet{zanella}.
\begin{table}
	\centering
	\caption[The dynamical parameters of the sources]{Concentration of the NFW dark-matter haloes, fixed parameters of the dynamical models presented in Section \ref{sec:src_dyn}.}
	\label{tab:c}
	\begin{tabular}{cc} 
		\hline
		\noalign{\smallskip}
		Name  & $c$\\
		\noalign{\smallskip}
		\hline
		\noalign{\smallskip}
		 SPT0113-46 & 3.04\\
		 SPT0345-47 & 2.97\\
		 SPT0441-46  & 2.77\\
		 SPT2146-55 & 2.67 \\
		SPT2132-58 &  2.45\\
    \noalign{\smallskip}
    \hline
    \noalign{\medskip}
	\end{tabular}
\end{table}

\begin{table}
	\centering
	\caption{Intervals of the prior distributions in the rotation curve decomposition (Section \ref{sec:src_dyn}).}
	\label{tab:prior}
	\begin{tabular}{cc} 
		\hline
		\noalign{\smallskip}
		Parameter & Prior \\
	    \noalign{\smallskip}
		\hline
		\noalign{\smallskip}
		 $M_{\mathrm{star}}$ & [$10^7, 10^{11}$] $M_{\odot}$ \\
		 $R_{\mathrm{e}}$ & [0.04, 4.0] kpc\\
		 $n$ & [0.5, 10]\\
		 $\alpha_{\mathrm{[CII]}}$ & [3.8, 238.0]$M_{\odot}/L_{\odot}$\\
		 $M_{\mathrm{DM}}$ & [$10^{10}, 10^{13}$] $M_{\odot}$\\
    \noalign{\smallskip}
    \hline
    \noalign{\medskip}
	\end{tabular}
\end{table}

\begin{figure*}
    \centering
    \includegraphics[width=0.8\textwidth]{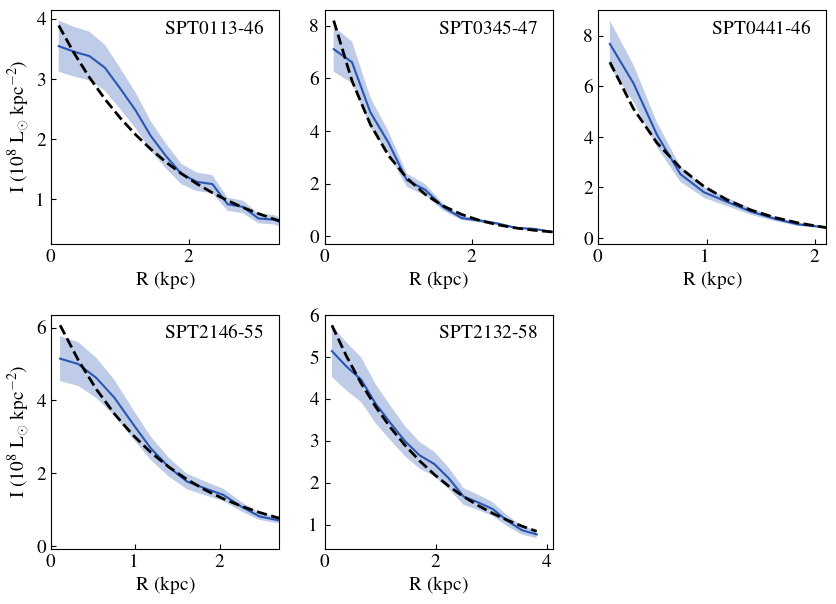}
    \caption{The blue solid lines and the shaded areas show the [CII] profiles and the corresponding uncertainties. These profiles are obtained dividing the zeroth-moment map of the reconstructed sources into rings (with centers, $PA$ and $i$ defined by the values of the kinematic model, Table \ref{tab:tab_kin}) and calculating the surface densities at a certain radius as azimuthal averages inside that ring. The black dashed lines show the best-fit exponential profiles.}
    \label{fig:exp}
\end{figure*}

\begin{figure*}
    \centering
    \includegraphics[width=0.9\textwidth]{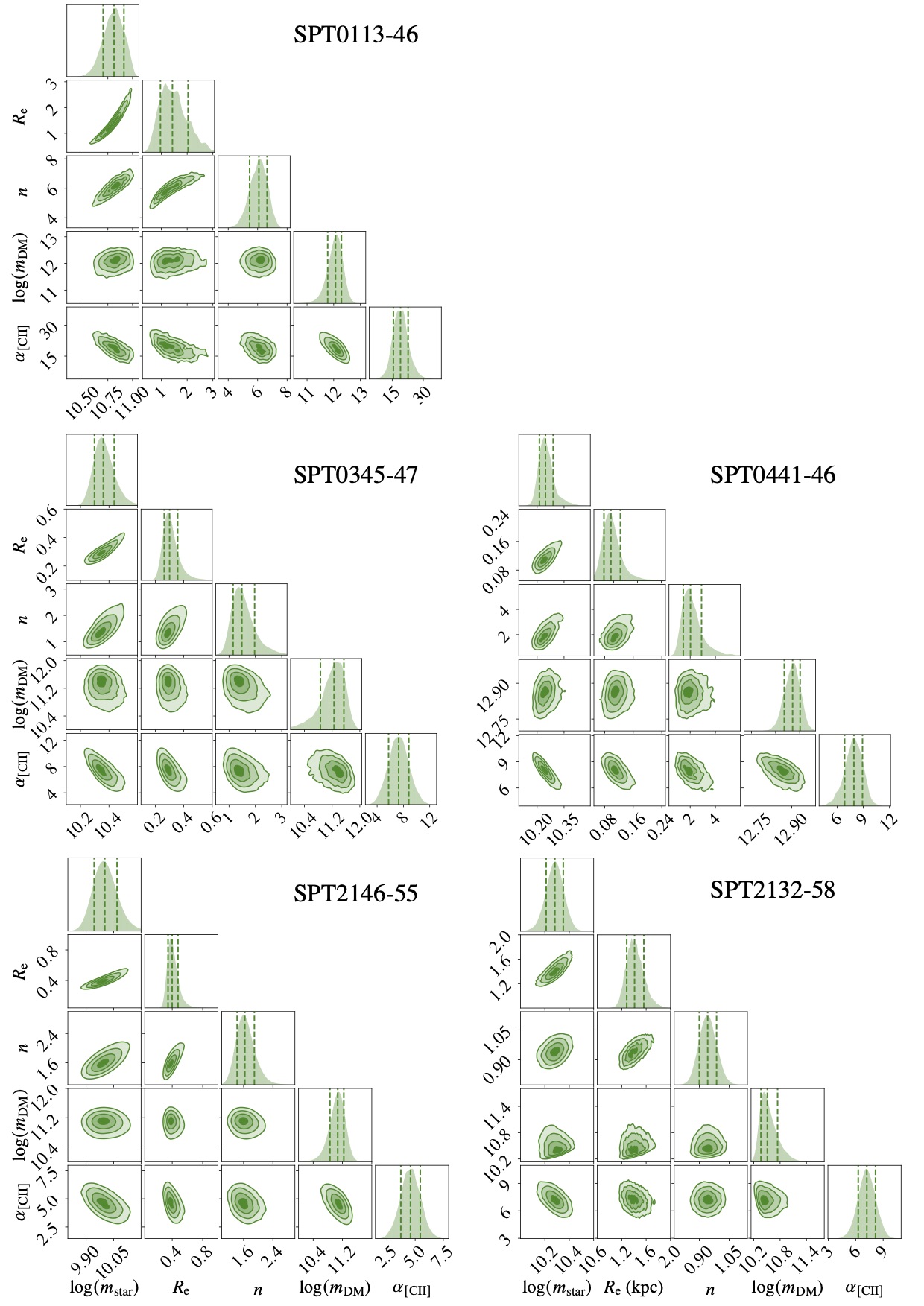}
    \caption{Posterior distributions of the dynamical parameters for the five galaxies studied in this paper, as indicated in the legend. The dashed lines in the 1D histograms show the 16th, 50th and 84th percentiles (see Table \ref{tab:dynamic1}). In each panel $m_{\mathrm{star}} = M_{\mathrm{star}}/M_{\odot}$ and $m_{\mathrm{star}} = M_{\mathrm{DM}}$/M$_{\odot}$. The units of  $R_{\mathrm{e}}$ and $\alpha_{[\mathrm{CII}]}$ are kpc and   M$_{\odot}$/L$_{\odot}$, respectively.}
    \label{fig:corner}
\end{figure*}


\bsp	
\label{lastpage}
\end{document}